\documentclass{aa}
\usepackage{txfonts}
\usepackage{amsmath}
\usepackage{color}
\usepackage{xcolor}
\usepackage[normalem]{ulem}
\usepackage{multicol}
\usepackage[colorlinks=true,citecolor=blue,linkcolor=blue]{hyperref}
\usepackage{textcomp}
\usepackage{subcaption}
\usepackage{graphicx}
\usepackage{float}
\usepackage{placeins}

\begin{document}

\title{Deep study of A399-401: Application of a wide-field facet calibration
\thanks{All the reduced images in this paper are only available at the CDS via anonymous ftp to 
\url{cdsarc.u-strasbg.fr} (\url{130.79.128.5}) or via \url{http://cdsweb.u-strasbg.fr/cgi-bin/qcat?J/A+A/}}}

\author{
J.M.G.H.J. de Jong\inst{\ref{inst:Leiden}} \and 
R.J. van Weeren\inst{\ref{inst:Leiden}} \and 
A. Botteon\inst{\ref{inst:Leiden},\ref{inst:UniBo},\ref{inst:inaf}} \and 
J.B.R. Oonk\inst{\ref{inst:Leiden},\ref{inst:surf},\ref{inst:Astron}} \and 
G. Brunetti\inst{\ref{inst:inaf}} \and 
T.W. Shimwell\inst{\ref{inst:Leiden},\ref{inst:Astron}} \and 
R. Cassano\inst{\ref{inst:inaf}} \and 
H.J.A. R\"ottgering\inst{\ref{inst:Leiden}} \and 
C. Tasse\inst{\ref{inst:Paris},\ref{inst:rhodes}}
}

\institute{
Leiden Observatory, Leiden University, PO Box 9513, 2300 RA Leiden, The Netherlands\relax\label{inst:Leiden} \and 
Dipartimento di Fisica e Astronomia, Universit\`a degli Studi di Bologna, via P. Gobetti 93/2, 40129 Bologna, Italy\relax\label{inst:UniBo} \and 
INAF - Istituto di Radioastronomia, via P. Gobetti 101, 40129, Bologna, Italy\relax\label{inst:inaf} \and 
SURF/SURFsara, Science Park 140, 1098 XG Amsterdam, The Netherlands\relax\label{inst:surf} \and 
ASTRON, The Netherlands Institute for Radio Astronomy, Postbus 2, 7990 AA Dwingeloo, The Netherlands\relax\label{inst:Astron} \and 
GEPI \& USN, Observatoire de Paris, Université PSL, CNRS, 5 Place Jules Janssen, 92190 Meudon, France\relax\label{inst:Paris} \and 
Department of Physics \& Electronics, Rhodes University, PO Box 94, Grahamstown, 6140, South Africa\relax\label{inst:rhodes}
}

\date{Received XXX 2022 / Accepted XXX 2022}

\begin{abstract} 
    {
        Diffuse synchrotron emission pervades numerous galaxy clusters, indicating the existence of cosmic rays and magnetic fields throughout the intra-cluster medium. It is general consensus that this emission is generated by shocks and turbulence that are activated during cluster merger events and cause a (re-)acceleration of particles to highly relativistic energies. Similar emission has recently been detected in megaparsec-scale filaments connecting pairs of premerging clusters. These instances are the first in which diffuse emission has been found outside of the main cluster regions.
    }
    {
        We aim to examine the particle acceleration mechanism in the megaparsec-scale bridge between Abell 399 and Abell 401 and assess in particular whether the synchrotron emission originates from first- or second-order Fermi reacceleration. We also consider the possible influence of active galactic nuclei (AGNs).
    }
    {
        To examine the diffuse emission and the AGNs in Abell 399 and Abell 401, we used deep (\textasciitilde 40 hours) LOw-Frequency ARray (LOFAR) observations with an improved direction-dependent calibration to produce radio images at 144 MHz with a sensitivity of $\sigma=79$ $\mu$Jy beam$^{-1}$ at a $5.9\arcsec\times 10.5\arcsec$ resolution. Using a point-to-point analysis, we searched for a correlation between the radio and X-ray brightness from which we would be able to constrain the particle reacceleration mechanism.
    }
    {
        Our radio images show the radio bridge between the radio halos at high significance.
        We find a trend between the radio and X-ray emission in the bridge. We also measured the correlation between the radio and X-ray emission in the radio halos and find a strong correlation for Abell 401 and a weaker correlation for Abell 399. On the other hand, we measure a strong correlation for the radio halo extension from A399 in the northwest direction. With our deep images, we also find evidence for AGN particle injection and reenergized fossil plasma in the radio bridge and halos.
    }
    {
        We argue that second-order Fermi reacceleration is currently the most favored process to explain the radio bridge. In addition, we find indications for a scenario in which past AGN particle injection might introduce significant scatter in the relation between the radio and X-ray emission in the bridge, but may also supply the fossil plasma needed for in situ reacceleration. The results for Abell 401 are also clearly consistent with a second-order Fermi reacceleration model. The relation between the thermal and nonthermal components in the radio halo in Abell 399 is affected by a recent merger. However, a strong correlation toward its northwest extension and the steep spectrum in the radio halo support an origin of the radio emission in a second-order Fermi reacceleration model as well. The evidence that we find for reenergized fossil plasma near Abell 399 and in the radio bridge supports the reacceleration of the fossil plasma scenario.
    }
    \keywords{galaxy clusters; cluster mergers; non-thermal radiation; intracluster medium; A399-401; radio bridges; reacceleration; radio halos}
    \maketitle
\end{abstract}

\section{Introduction}\label{sec: introduction}

Structures in our Universe are growing hierarchically, with smaller systems merging to form larger structures. The largest gravitationally bound structures are galaxy clusters, and when these merge with each other, \textasciitilde\,$10^{64}$ erg is released into the intracluster medium (ICM) on timescales of billions of years \citep{markevitch2007, hoeft2008}. The ICM is a diluted plasma that permeates the cluster volume and primarily emits thermal bremsstrahlung at X-ray wavelengths.
Synchrotron radio emission has been observed in numerous clusters \citep[see][for a recent review]{vanweeren2019}. The presence of this emission indicates the existence of cosmic rays and magnetic fields in the ICM. The general consensus is that shocks and turbulence, generated during cluster merger events, cause the (re)acceleration of particles to highly relativistic energies \citep{brunetti2014}. Recently, diffuse radio emission has also been detected between pairs of clusters at megaparsec (Mpc)-scales \citep{govoni2019, botteon2020}. These so-called radio bridges might trace regions in which the gas is compressed during the initial phase (i.e., the premerger phase) of the collision between massive galaxy clusters. Radio observations of cluster bridges open new windows for studying the acceleration of cosmic rays in environments with a density that is lower than typical in clusters \citep{brunetti2020}. The detection of radio bridges also brings us closer to the detection and study of plasma conditions in the densest phase of the so-called warm-hot intergalactic medium (WHIM) \citep{vazza2019}. However, because only a few bridges are known and only very few theory papers have been published about their possible origin, the investigation of the origin of the magnetic fields and cosmic rays in the radio bridges is still in an initial stage in these low-density environments.

Radio bridges associated with the premerging clusters Abell 1758N and Abell 1758S (A1758) at $z=0.279$ \citep{botteon2018, botteon2020} and the premerging clusters Abell 399 and Abell 401 (A399-401) at $z=0.072$ \citep{govoni2019} have been recently discovered. These radio bridges are between two comparable systems and were discovered with LOw Frequency ARray (LOFAR) observations at 144 MHz. Follow-up studies at different frequencies have been performed recently \citep{botteon2020, nunhokee2021}. Moreover, Abell 1430 might have a radio bridge between two merging clusters (a main cluster and subcluster), but this has not been fully confirmed \citep{hoeft2021}. Other types of radio bridges have been discovered between the Coma cluster and the NGC4839 group ($z\thickapprox 0.0231$) at 346 MHz \citep{kim1989} and 144 MHz \citep{bonafede2021} and between the cluster Abell 3562 and the radio source J 1332-3146a in the group SC 1329-313 in the Shapley supercluster ($z\thickapprox 0.048$) at GHz frequencies \citep{venturi2022}. 
Of all the bridges between premerging clusters, A399-401 has been most frequently and deeply studied with X-ray observations and Sunyaev-Zeldovich (SZ) effect measurements \citep{fujita1996, fabian1997, markevitch1998, sakelliou2004, fujita2008, murgia2010, planck2013, akamatsu2017, bonjean2018, hincks2022}. It has already been known for a while that A401 has a radio halo \citep{harris1980, roland1981, bacchi2003}, but \cite{murgia2010} identified a radio halo in A399 as well, which made A399-401 the first detected double radio-halo system. The detection of these radio halos suggests that the clusters themselves are also undergoing their own mergers.

Because of energy losses, relativistic electrons can only travel up to sub-Mpc distances at 140 MHz \citep{jaffe1977} in their lifetime. These age constraints mean that
particles must be generated in situ to explain how diffuse radio emission can originate on Mpc scales in the A399-401 bridge.
 \cite{govoni2019} proposed that radio bridges may result from first-order Fermi (Fermi-I) reacceleration of a volume-filling population of fossil relativistic electrons by weak, $\mathcal{M}\leq 2-3$, shocks under favorable projection effects. 
Alternatively, \cite{brunetti2020} suggested that the synchrotron emission from the radio bridge could be a result of second-order Fermi (Fermi-II) reacceleration, where turbulence plays a major role by amplifying magnetic fields and reaccelerating particles. In this case, preexisting relativistic particles and magnetic fields interact with the turbulence, which reenergizes them in the region between the two clusters. \cite{nunhokee2021} recently constrained a steep spectrum ($\alpha>1.5$) supporting a turbulent Fermi-II reacceleration origin.
\cite{botteon2020} found a trend between the radio and X-ray emission in the bridge A1758 by studying the spatial correlation between the two emission components. This suggests that the radio and X-ray emissions are generated in comparable volumes, which supports turbulent reacceleration. Strong spatial correlations have been observed for radio halos as well \citep{govoni2001, feretti2001, giacintucci2005, rajpurohit2018, rajpurohit2021, botteon2020b, ignesti2020, biava2021, duchesne2021, bonafede2021}, where Fermi-II reacceleration in most cases been understood to be the most relevant particle acceleration process for giant radio halos \citep{vanweeren2019}.

The goal of this paper is to study the morphology and origin of the synchrotron emission in A399-401 in more detail. We use new, deep radio data for this aim and an improved direction-dependent (DD) calibration method. With the new radio map, we study the diffuse emission from the radio halos and radio bridge in greater depth and investigate new features related to the origin of the reaccelerated particles. Additionally, we correlate our new radio surface brightness map with an X-ray surface brightness map from A399-401 as a tool for inferring the mechanism behind the reacceleration of electrons in the diffuse emission from the bridge and radio halos.

We start by describing the data and the data reduction method in Section \ref{section:datacalibration}. The radio images are discussed in Section \ref{section:scienceresults}. In Section \ref{sec:thermalnonthermal} we consider the relation between the radio and X-ray emission. All the results are discussed in Section \ref{section:discussion}. Finally, we conclude our work in Section \ref{section:summary}.

We use a $\Lambda$CDM cosmology model with $H_0=70$~km~s$^{-1}$~Mpc$^{-1}$, $\Omega_{m}=0.3$, and $\Omega_{\Lambda}=0.7$. The images in this paper are made in the J2000 coordinate system.

\section{Observations and data reduction}\label{section:datacalibration}

In this section, we describe the radio data that we used and how we calibrated them to arrive at a final image that we used for our science. Because we wish to relate the radio and X-ray emission, we also reduced X-ray observations.

\subsection{Data}\label{section:data}

For the study of A399-A401, we used 6$\times$8 hours of observations from LOFAR \citep{haarlem2013} from project \texttt{LC9\_015} (PI: van Weeren). The observation IDs and observation dates are given in Table \ref{table:ms}. Every observation has a pointing center at a right ascension of 02h 58m 21s and a declination of +13\textdegree 17' 10$\arcsec$ (J2000 equinox). The data cover the frequency range from 120--168\,MHz and were observed only with high-band antennas. We only used the  stations located in the Netherlands. L626692 used 60 stations, while the other observations all used 62 stations.

\begin{table}
 \caption{Measurement set names with observation dates.}
 \label{table:ms}
\begin{center}
\begin{tabular}{|l | c | c ||}
 \hline
 \bf{Name} & \bf{Date} \\ [0.5ex] 
 \hline\hline
L626678  & 7/12/2017 \\ 
L626692  & 30/11/2017 \\ 
L626706  & 16/11/2017 \\ 
L632229  & 13/12/2017 \\ 
L632511  & 27/12/2017 \\
L632525  & 20/12/2017 \\
\hline
\end{tabular}
\end{center}
\end{table}

During the testing of our calibration method (further discussed in Section \ref{section:calibrationmethod}), we decided to flag the last 1h 20m from all the observations, which leaves a total of 40 hours of observation time. This was necessary because the calibration solutions started to diverge in this part, which was most likely caused by a low elevation that decreases the sensitivity and means a thicker ionosphere to look through. Moreover, we also manually flagged a few sub-bands between 126 and 128 MHz because they were contaminated by radio frequency interference (RFI). The central frequency of our data is 144 MHz.

\subsection{Calibration}\label{section:calibrationmethod}

The two main parts in the calibration of LOFAR data are direction-independent (DI) and DD calibration. The DI calibration follows the LOFAR Two-metre Sky Survey (LoTSS), where \texttt{Prefactor} version 3 with the Default Preprocessing Pipeline (\texttt{DPPP}) is used \citep{weeren2016, williams2016, gasperin2019}.\footnote{\url{https://github.com/lofar-astron/prefactor}} This includes RFI flagging \citep{offringa2012}, bandpass corrections, removing data that are affected by bright off-axis sources, clock-TEC separation, polarization alignment, ionospheric rotation measure (RM) corrections, and calibrating against a sky model from external radio surveys.
The implementation of the automated DI calibration pipeline itself is discussed in \cite{mechev2017}. The DI calibration has been left untouched because the main remaining calibration issues were coming from DD solutions near bright sources around the clusters.

The DD effects are caused by ionospheric effects and imperfect beam models, which can be corrected with Jones matrices \citep{hamaker1996, shimwell2019}, derived from the visibilities.
Over time, several DD correction algorithms have been developed, for instance, \texttt{SPAM} \citep{intema2009}, \texttt{Sagecal} \citep{sage2011}, or facet calibration \citep{weeren2016}. For LoTSS, the \texttt{DDF-Pipeline} has been developed by the LOFAR Surveys Key Science Project, which is based on \texttt{KillMS} to derive the Jones matrices and to apply the solutions during the imaging of the entire field of view (FoV) with \texttt{DDFacet} \citep{tasse2014, tasse2018, tasse2021, smirnov2015, shimwell2019}.\footnote{\url{https://github.com/mhardcastle/ddf-pipeline}}
With these DD calibrations, LoTSS 8-hour observations reach 6$\arcsec$ angular resolution and a typical median sensitivity of $\sigma\approx 80$ $\mu$Jy beam$^{-1}$ over the entire LoTSS-DR2 area \citep{shimwell2022}.

Although the output from the automated \texttt{DDF-Pipeline} is sufficient to do most science with, there is still room for improvements, especially for targets with large angular extent. We therefore decided to further enhance the data reduction for this specific field. Our goal was to correct for artifacts around bright sources near A399 and A401, which determine how deep we can look into the substructure from the diffuse radio bridge and how much we can detect from the radio halos. 
One of the main parts to improve is the selection of specific directions to derive and apply DD calibrations. The \texttt{DDF-Pipeline} makes corrections in a user-specified number of directions (45 is used for LoTSS standard processing). The number of directions constrains the facet layout and the final calibrated image because it is assumed that the DD calibration solutions are constant throughout a facet. Because the directions are determined in an automated way, these layouts are not always optimal. This motivated us to use the recalibration method described in \cite{vanweeren2021}. This calibration method has already been successfully used in numerous other works \citep{botteon2019, botteon2020b, botteon2020,botteon2021n, botteon2022, cassano2019, hardcastle2019n, osinga2021, hoang2021}. Because A399-401 covers a large area, however, we need to apply this method for many directions, which required  including additional steps. In the following sections, we describe the recipe for a single direction ($N=1$) from \cite{vanweeren2021}, followed by a explanation how we upgraded this to several directions ($N>1$), and applied it to A399-401. 

\begin{figure*}[p]
    \begin{center}
        \includegraphics[width=0.67\textwidth]{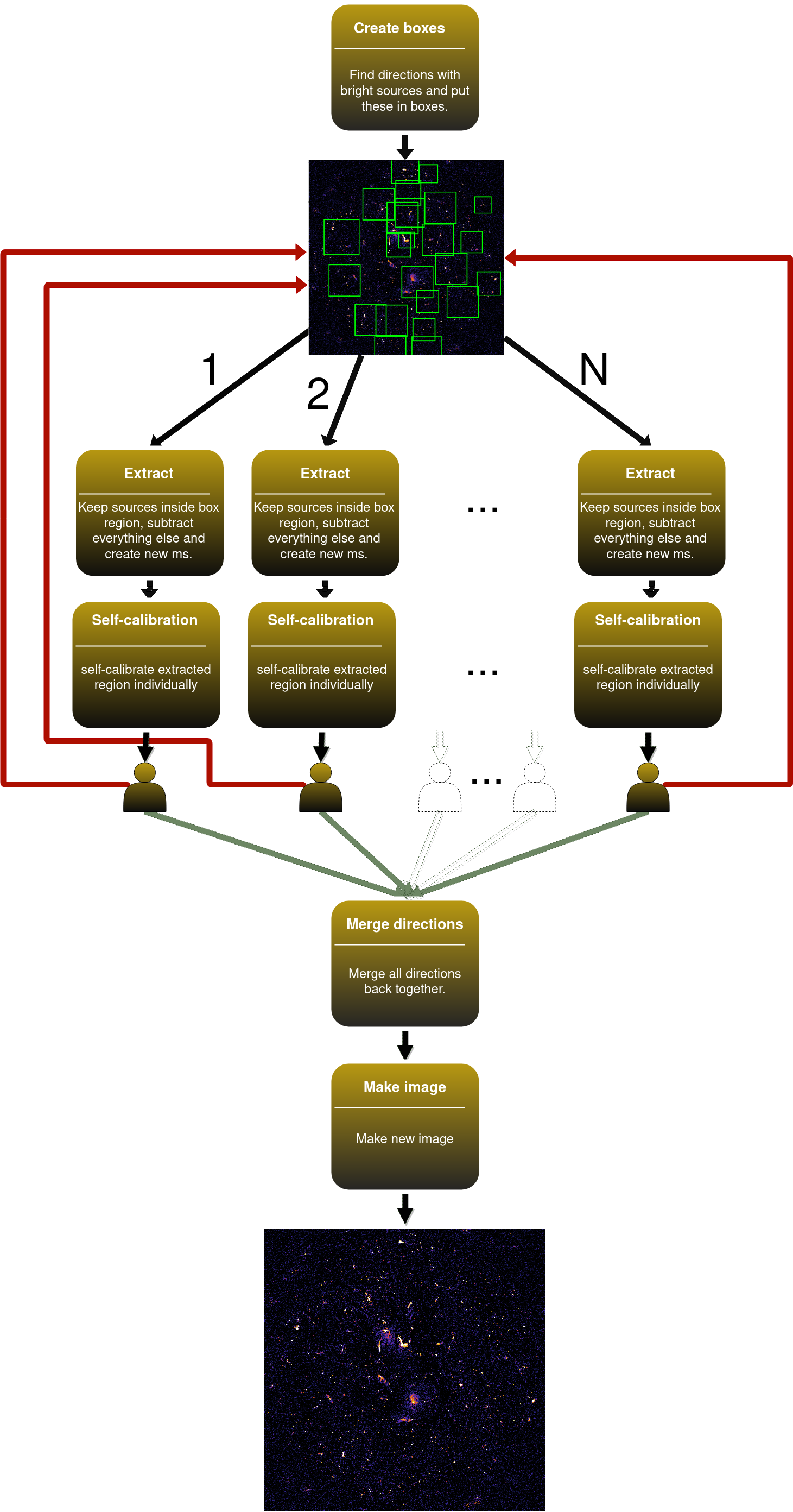}
    \end{center}
    \caption{Flowchart showing every major step from the recalibration recipe described in Sections \ref{n=1} and \ref{n>1}.}
    \label{fig:flowchart}
\end{figure*}

\subsubsection{N=1}\label{n=1}
\begin{figure}
    \begin{center}
        \includegraphics[width=0.48\textwidth]{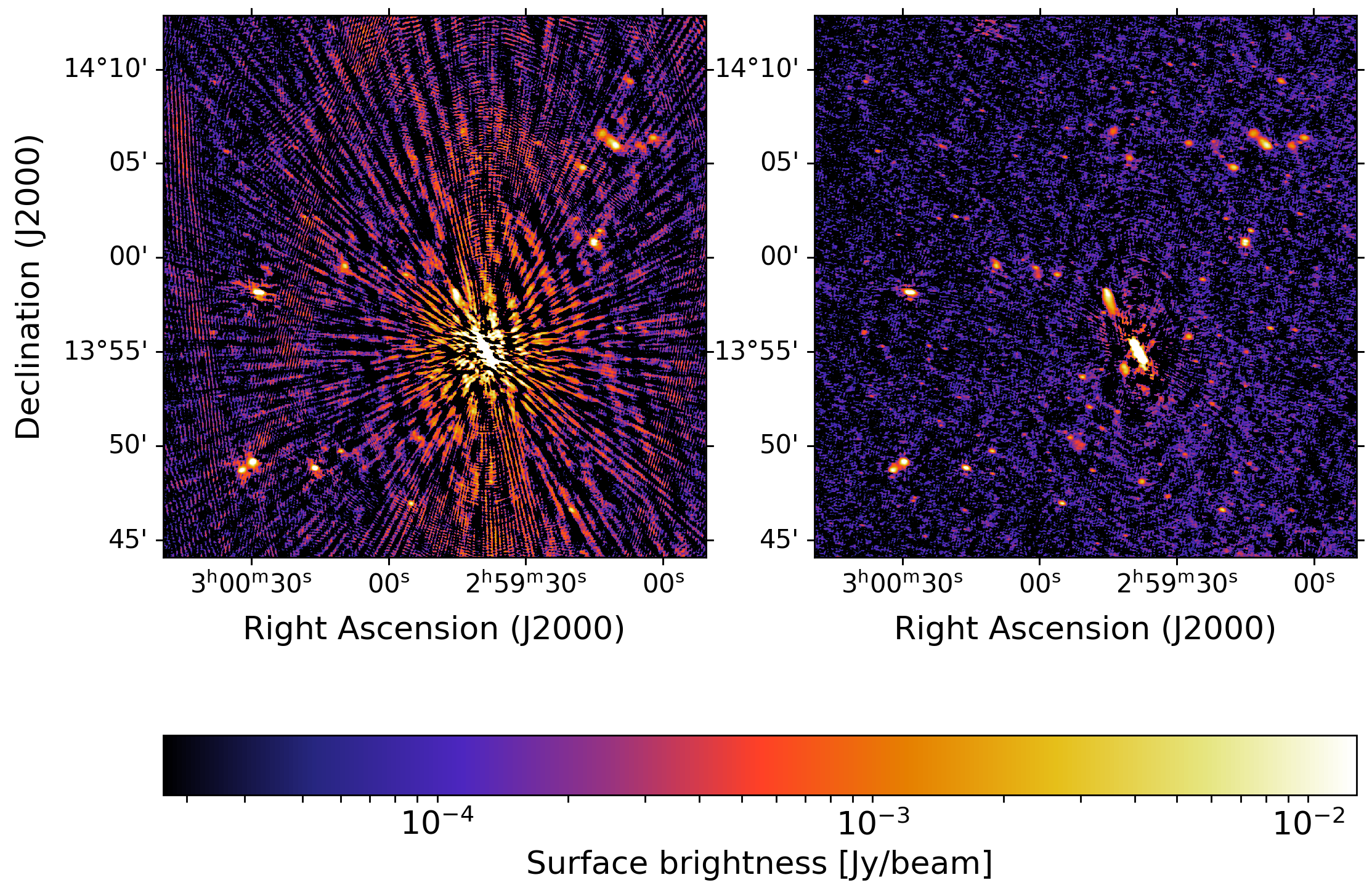}
    \end{center}
    \caption{Self-calibration of an individual box. \textit{Left panel}: Image of an extracted box before self-calibration and with the DI solutions alone. \textit{Right panel}: Image of the same region after eight self-calibration cycles. Visible artifacts around the source disappear while correcting for phase effects. These images are made with \texttt{WSClean} version 3 \citep{wsclean2014}.}
    \label{fig:selfcal}
\end{figure}

First, we made a square box region with the \texttt{DS9} software around a bright compact source \citep{ds9}. In this area, self-calibration was applied and DD effects were corrected for. All sources outside of this box were subtracted from the visibilities with the DD solutions and sky model from the \texttt{DDF-Pipeline} \citep[extraction step, see][]{vanweeren2021}. Ideally, the box had sides with a size between 0.25\textdegree\,and 0.4\textdegree. Box sizes need to be large enough for the flux density to be high enough so that diverging solutions are avoided in the self-calibration, whereas to improve upon the \texttt{DDF-pipeline,} the boxes need to be smaller than the facets used there because both assume constant calibration solutions across the facet. After the extraction, we phase-shifted the \textit{uv}-data to the center of the box and averaged the time and frequency to 16s and 0.39 MHz to reduce the data size by a factor of 8, which is sufficient for smearing purposes and does not lead to ionospheric calibration problems. With \texttt{Dysco,} we further compressed the data volume \citep{offringa2016compression}. Then followed several rounds of self-calibration on the extracted box (self-calibration step). The starting point were the DI calibrations from the \texttt{DDF-Pipeline}. In all self-calibration rounds, we performed several so-called \texttt{tecandphase} calibrations with \texttt{DPPP} \citep{dppp2018} to solve for the total electron content (TEC). We achieved this with solution intervals between 16s and 48s, followed by Stokes I gain calibrations with preapplied \texttt{tecandphase} solutions and solution intervals between 16 and 48 minutes along the time axis and solution intervals between 2 MHz and 6 MHz along the frequency axis. These solution intervals were automatically determined for each box region based on the amount of apparent compact source flux, as this differs per box.
After all rounds of self-calibration, we created the final image of the facet. This was done with \texttt{WSClean} \citep{wsclean2014} or the \texttt{DDFacet} imager \citep{tasse2018}. See Figure \ref{fig:selfcal} for an example of the result after eight self-calibration cycles, imaged with \texttt{WSClean}.

\subsubsection{N>1}\label{n>1}
We scaled the method in Section \ref{n=1} in order to allow the use of an arbitrary number of box regions ($N>1$). In every box, we included at least one bright source. To limit the manual steps and to save time, we implemented an automated box-region generator. We found bright sources by scanning for all pixels where the surface brightness is higher than 70 mJy beam$^{-1}$ in the image with a resolution of $6\arcsec$  from LoTSS that we wished to recalibrate. This pixel value was chosen because we found it to correspond to the approximate flux limit for a stable calibration for each box. We started with the brightest source and placed an initial box with a size of 0.4\textdegree $\times$0.4\textdegree. 
Because a smaller box size speeds up the self-calibration but enough flux is necessary, the algorithm optimizes the size for each box with final sizes between 0.25\textdegree\, and 0.4\textdegree\  while at the same time finding an optimal box center not farther than 0.2\textdegree\, from the initial position. After the full box layout was optimized, we manually further fine-tuned the result to obtain the optimal composition. This was deemed to be necessary for difficult cases in which many compact bright sources are near each other, which makes it difficult for the algorithm to decide whether to include them in different or the same boxes. When the box layout was approved (see, e.g., the left panel in Figure \ref{fig:layout}), we followed all the same steps as in the $N=1$ case for every individual box. Boxes may overlap because the solutions that are applied to a part of the sky are taken from the closest box center. In some cases, this overlap is necessary to obtain enough flux for the self-calibration. When all the individual self-calibrations were completed, we validated the quality of every set of solutions, such that no corrupt or diverging solutions were later applied in the imaging step. After the calibration, the solutions were merged into a single \texttt{HDF5} solution file per observation.\footnote{\url{https://github.com/jurjen93/lofar_helpers/blob/master/h5_merger.py}}
The box layout can be mapped to a facet layout, as we show in Figure \ref{fig:layout}. These facets represent the final solution area with solutions from the closest box to each pixel in the image. The solutions from these facets were applied in the final imaging step. To do this, it is only possible to use an imager that supports facets (in our case, the facet mode of \texttt{WSClean} version 3). All the main steps from box selection until imaging are summarized in Figure \ref{fig:flowchart}.

\subsubsection{Facet calibration for A399-401}
\begin{figure*}[ht]
  \centering
  \includegraphics[width=0.48\textwidth]{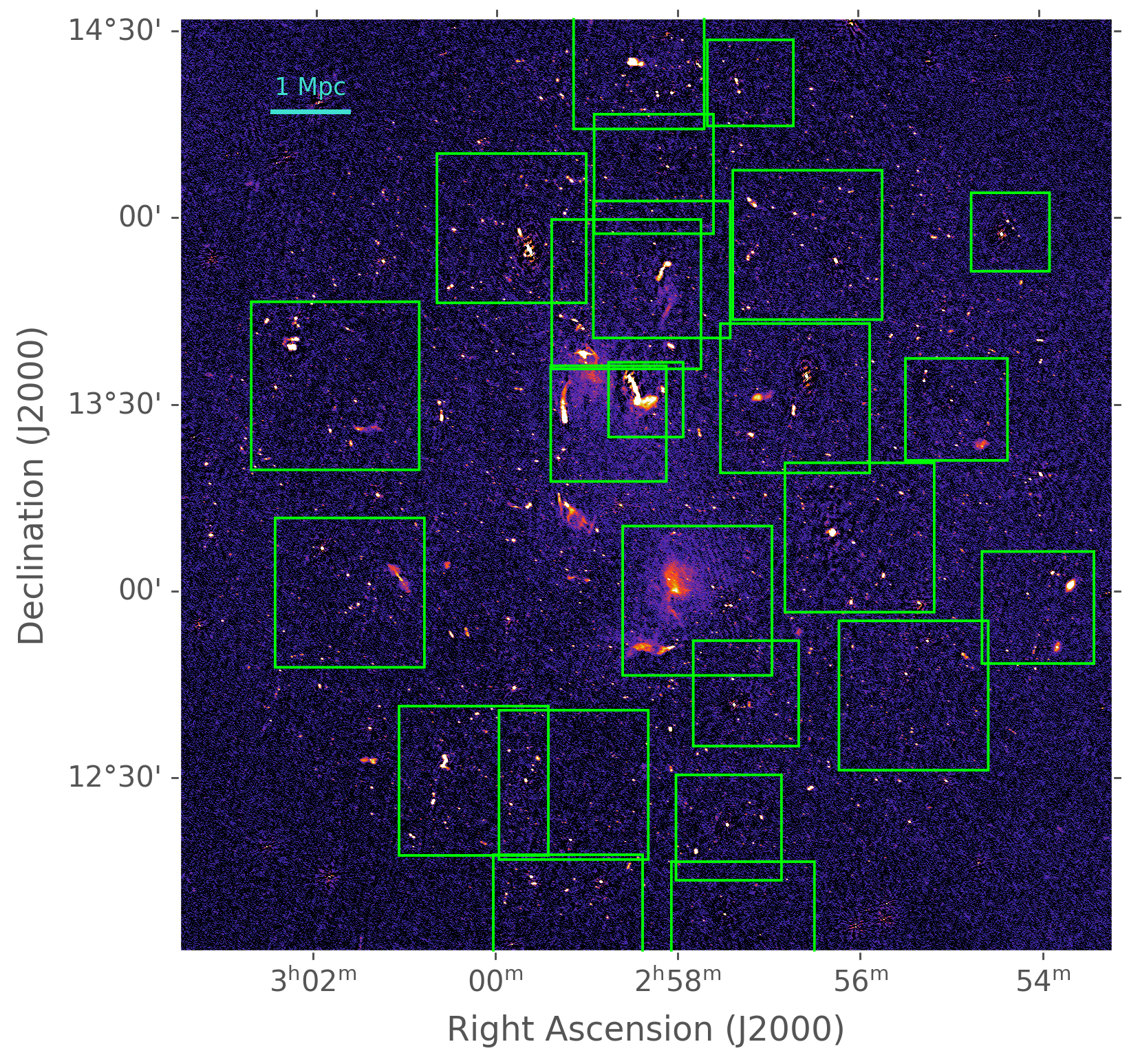}  
  \centering
  \includegraphics[width=0.48\textwidth]{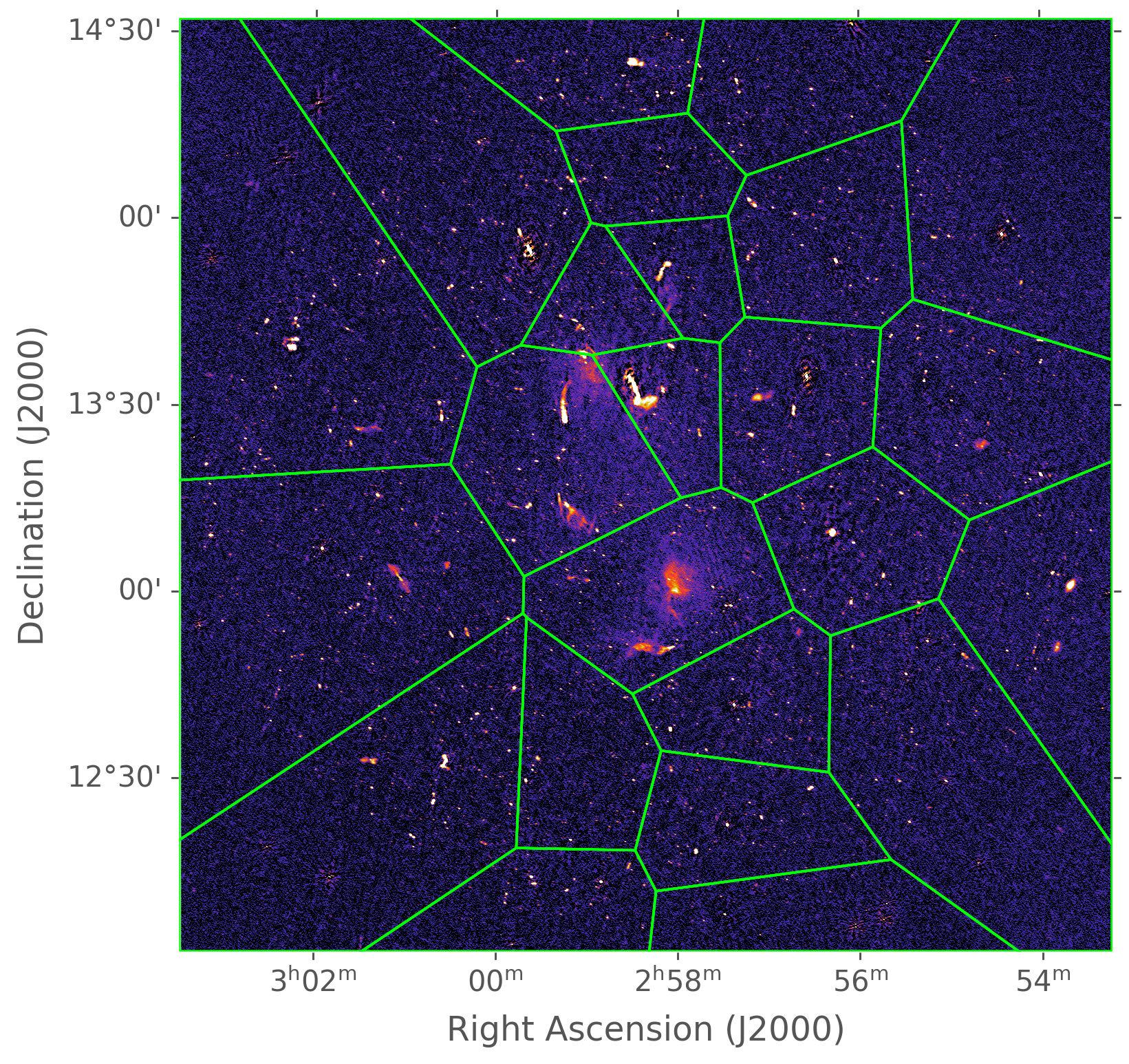}  
\caption{Mapping from box layout to facet layout. \textit{Left panel}: Box layout for A399-401 within a radius of 0.6\textdegree~from the pointing center, where all individual self-calibration boxes are contoured in green. \textit{Right panel}: Facet layout corresponding to the box layout from the left panel. The size of 1\,Mpc is given in the top left corner for $z=0.072$.}
\label{fig:layout}
\end{figure*}

Using the method from Section \ref{n>1} for $N>1$, we recalibrated an area with a radius of 1.2\textdegree\, from the pointing center of A399-401 with N=24 boxes. 
This small region size was chosen to reduce the computational cost by a factor \textasciitilde 4 compared to recalibrating and imaging the full field of view of our pointing.\footnote{We needed 50336 CPU core hours for the recalibration (see Appendix \ref{appendix:spider}), which would have been \textasciitilde 189000 CPU core hours for the full field of view.} This choice does not affect the result of our main target of interest, which is in the center of the field and extends for \textasciitilde 0.5\textdegree. Everything outside this area was subtracted from the visibilities. In Figure \ref{fig:layout} the final box and facet layout for our field are shown. Every box corresponds to a different facet.

For A399-401 we added another reduction in the computational costs by using eight self-calibration cycles instead of the standard ten cycles for every box in \cite{vanweeren2021}, as we realized that the noise level in the last two rounds of self-calibration did not reduce or there were no improvements at all. Another speedup was added by applying an additional factor 4 of time and frequency averaging in the first five rounds and by returning to the unaveraged data for the last three rounds. This did not affect the final result in a noticeable way, as we obtained similar results with or without this additional averaging. As all extractions and self-calibrations can run independent of each other on different computing nodes (i.e., it is an embarrassingly parallel problem), parellelization accelerated the total processing time with a factor \textasciitilde\,20. In Appendix \ref{appendix:spider}, we provide details about the computational cost of our recalibration. 

After obtaining all self-calibration solutions, we had one merged \texttt{HDF5} solution file and measurement set for each observation. These were then used for the final imaging with the facet mode from \texttt{WSClean} version 3, using multiscale multifrequency deconvolution and Briggs weighting in facet mode \citep{briggs1995, wsclean2014, offringasmirnov}. 
\texttt{DDFacet}  also has a facet mode, but we chose \texttt{WSClean} because in our tests on this field, the deconvolution performs better for extended emission with this imager. 
Our final image had a resolution of $5.9\arcsec\times 10.5\arcsec$ , and we reached a sensitivity of $\sigma=79$ $\mu$Jy beam$^{-1}$ at 144 MHz. We also further tapered the visibilities to obtain an image at $24.6\arcsec\times27.1\arcsec$ with $\sigma=230$ $\mu$Jy beam$^{-1}$, and at $72.8\arcsec\times75.9\arcsec$ with $\sigma=809$ $\mu$Jy beam$^{-1}$. These lower-resolution images have a better surface brightness sensitivity and allow us to better recover the diffuse extended emission from the radio bridge.

\subsubsection{Advantages and disadvantages of recalibration}

We can compare our highest-resolution recalibrated radio map with the radio map produced by the standard \texttt{DDF-pipeline}, which is based on the same observations. This pipeline is also used for LoTSS. By visual inspection, we see fewer artifacts around bright compact sources, and the diffuse emission is better reconstructed in our radio map than in the \texttt{DDF} image. We quantified this by studying the dynamic range around these compact sources. For most cases, this also improved (by a factor \textasciitilde1.6). In Appendix \ref{ddfvsnew} we elaborate on this comparison. Overall, we can conclude that the recalibration method we used is a useful tool for calibrating a large area (larger than \textasciitilde 0.8\textdegree) in which calibration artifacts remain around bright sources after using the \texttt{DDF-Pipeline}. However, the high additional computational costs make it a very expensive method at present (see Appendix \ref{appendix:spider}). The flowchart from Figure \ref{fig:flowchart} is not a full working pipeline either, which makes the implementation not straightforward. These advantages and disadvantages need to be considered or optimized in future usage of this method.

\subsubsection{Removing compact sources}\label{section:compact}

Because we are interested in the diffuse radio emission from the A399-401 radio bridge and radio halos and aim in Section \ref{section:ptp} to compare this emission with an X-ray map, we also created additional images from which the contribution from discrete compact sources was removed.
As there is no perfect way to do this, we applied two different methods. Both have their advantages and disadvantages. 

In the first method, we start by obtaining a compact source model by making an image from which we remove the shortest baselines corresponding to a certain physical scale. This prevents extended emission from entering the model. Then, we subtract the clean components from this high-resolution model from the starting \textit{uv} data. With these new \textit{uv}-subtracted data, we can make an image that is tapered to a lower resolution of $72.8\arcsec\times75.9\arcsec$, where contribution from compact sources is subtracted and extended emission is enhanced.
We tried several baseline cuts corresponding to 200\,kpc, 300\,kpc, 400\,kpc, and 500\,kpc at the redshift of A399-401. Based on visually inspecting and comparing the final results with the original nonsubtracted image, we decided to use the 300\,kpc scale, as this gave the best balance between removing compact sources and having no noticeable impact on the diffuse emission. This corresponds to $216\arcsec$ and $943\lambda$. Although the \textit{uv}-subtract method succeeds in keeping the diffuse emission and removing most of the compact sources, there are often leftover sources mainly from extended AGNs, which can affect flux density measurements.

For the second method, we use the open map filter from \cite{rudnick2002} (R02 filter) to remove compact sources directly in image space. This method applies a sliding minimum filter, followed by a sliding maximum filter with the same kernel size on the image data. The R02 filter is sensitive to the noise and the kernel size.
This becomes more prominent when we filter in more diffuse areas with a low signal-to-noise ratio. On the other hand, this filter is very efficient in removing all compact sources smaller than the used kernel size. However, it does not remove compact sources larger than the kernel size and can leave residual emission from partially subtracted extended AGNs.
By experimenting with different settings, we decided to apply this filter on our $24.6\arcsec\times27.1\arcsec$ map with a kernel size of $42\arcsec$ (corresponding to 60\,kpc at the redshift of A399-401) and further smooth this to $72.8\arcsec\times75.9\arcsec$ to have the same resolution as the other radio map.

\subsubsection{X-ray data}

We retrieved archival \textit{XMM-Newton} observations of A399-401 from the Science Archive\footnote{\url{http://nxsa.esac.esa.int/nxsa-web}}. In particular, we made use of three pointings: 0112260101 (covering A399), 0112260301 (covering A401), and 0112260201 (covering the region between the two clusters). The European Photon Imaging Camera (EPIC) observations were processed with the \textit{XMM-Newton} Scientific Analysis System (SAS v16.1.0) and the Extended Source Analysis Software (ESAS). After filtering bad time intervals due to soft proton flares, we produced an EPIC mosaic image in the $0.5-2.0$ keV band combining the three ObsIDs. This was used to compare the X-ray and radio (from LOFAR) surface brightnesses of the observed emission. For a detailed analysis of the \textit{XMM-Newton} observations, we refer to \citet{sakelliou2004}.

\section{Results}\label{section:scienceresults}

\begin{figure*}[ht]
\centering
  \includegraphics[width=0.9\textwidth]{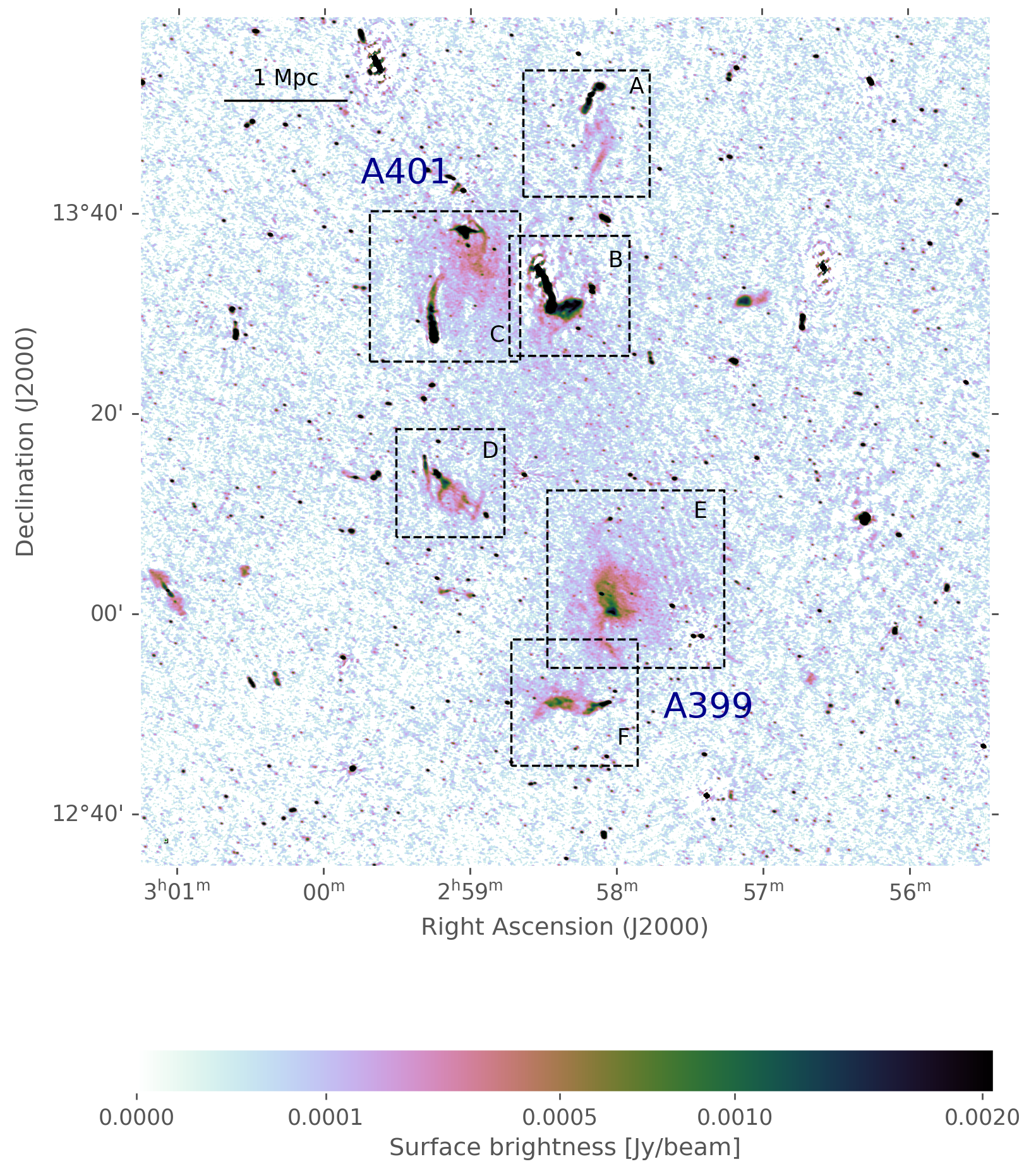}  
\caption{Radio map of A399-401 at a resolution of $5.9\arcsec\times 10.5\arcsec$ with $\sigma=79$ $\mu\text{Jy beam}^{-1}$. The black highlighted regions correspond to the zoomed images in Figure \ref{fig:subimages}.
 The beam size is given in the lower left corner, and the scale of 1\,Mpc at $z=0.072$ is given in the upper left corner. The square-root color scale of the images extends from 0 to $25\sigma$.}
\label{fig:6arc}
\end{figure*}

\begin{figure*}[ht]
\centering
  \includegraphics[width=0.48\textwidth]{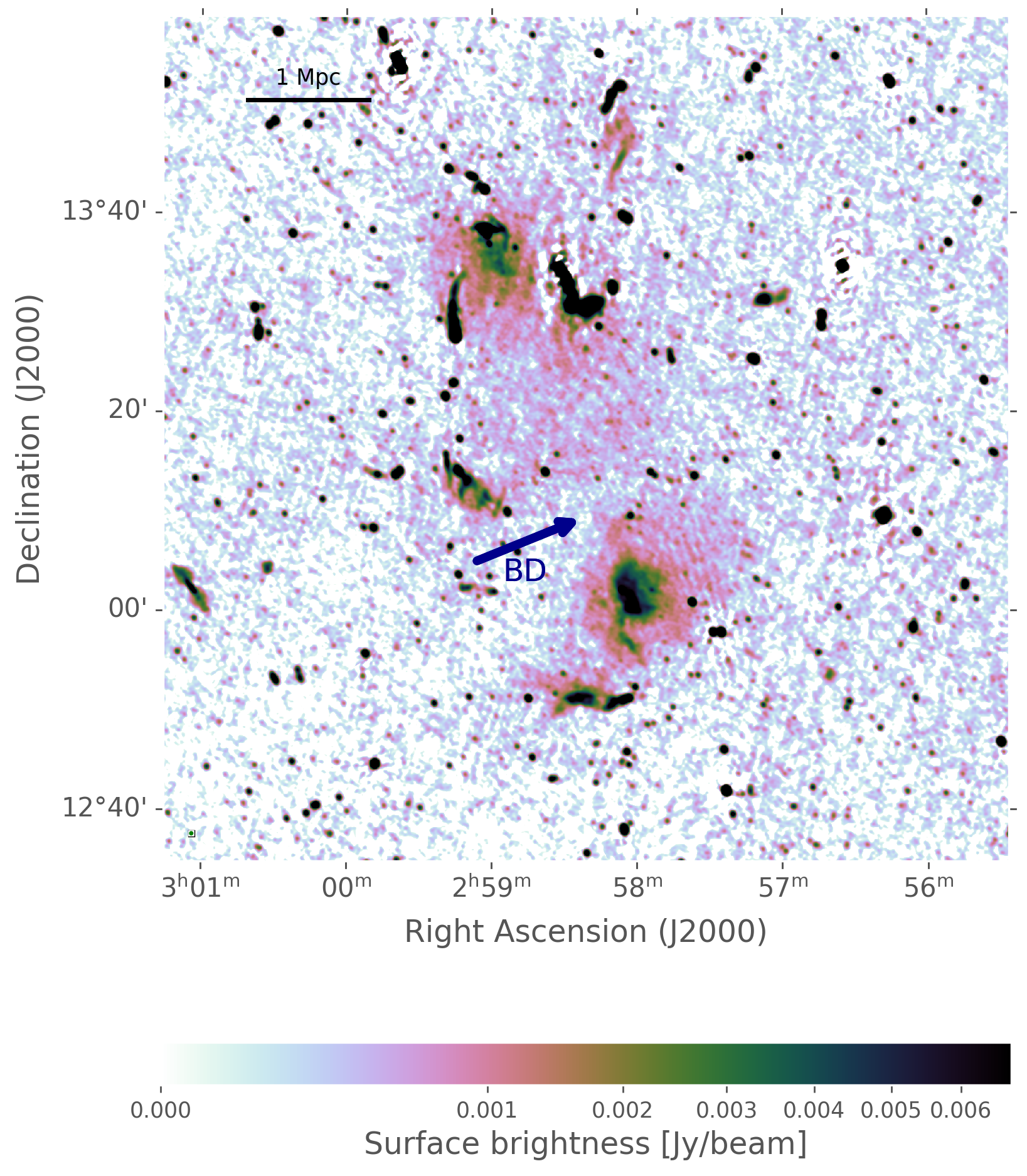}  
  \includegraphics[width=0.48\textwidth]{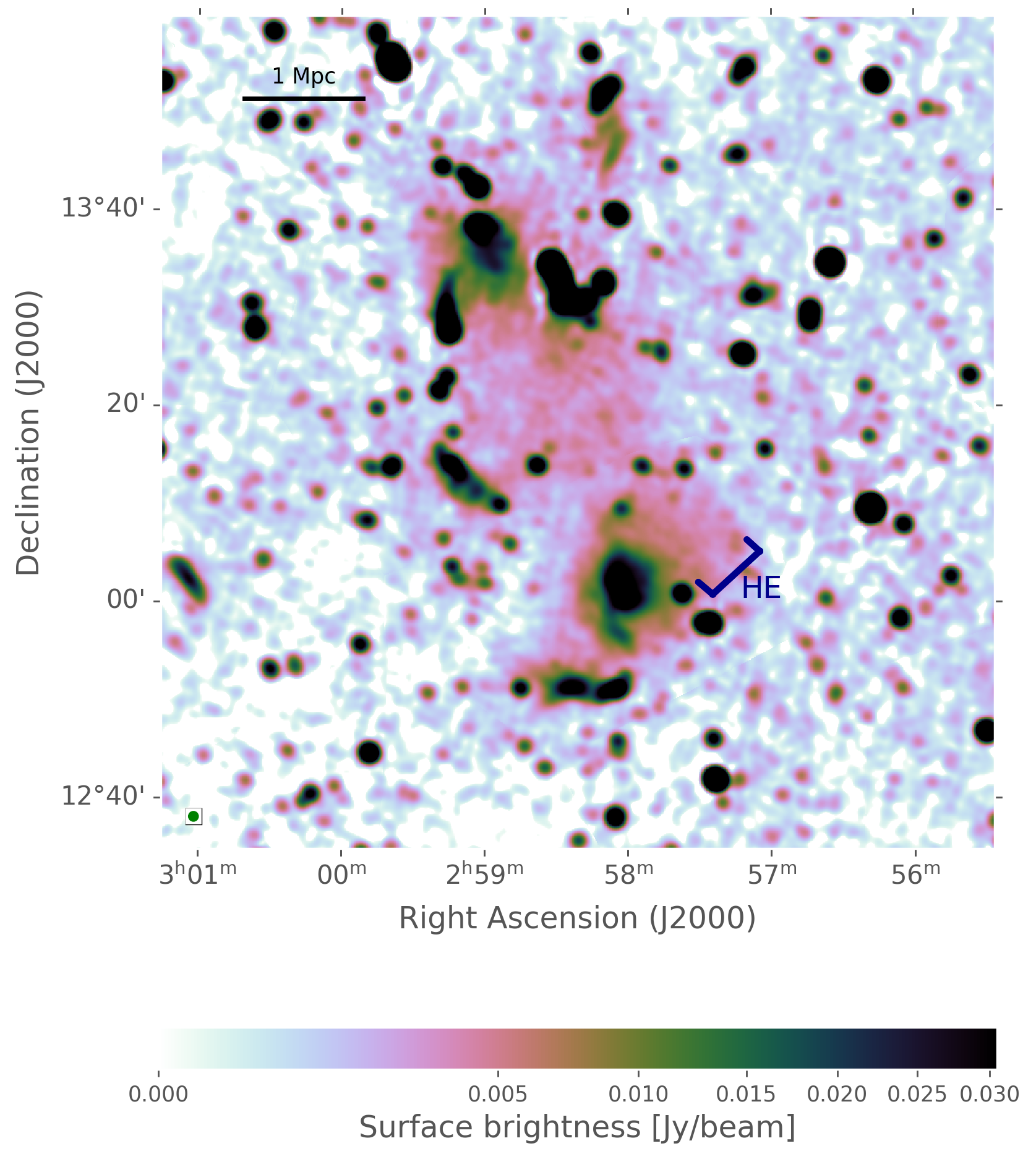}  
\caption{Radio maps of the A399-401 system at lower resolutions. \textit{Left panel}: Resolution of $24.6\arcsec\times27.1\arcsec$ with $\sigma=230$ $\mu$Jy beam$^{-1}$. \textit{Right panel}: Resolution $72.8\arcsec\times75.9\arcsec$ with $\sigma=809$ $\mu$Jy beam$^{-1}$. The dark blue brightness depression (BD) and the halo extension (HE) are described in the text. The beam size in all images is given in the lower left corner, and the scale of 1\,Mpc at $z=0.072$ is given in the upper left corner.  The square-root color scale
of the images extends from 0 to $25\sigma$.}
\label{fig:lowres}
\end{figure*}

\begin{figure*}[ht]
  \centering
  \includegraphics[width=0.48\textwidth]{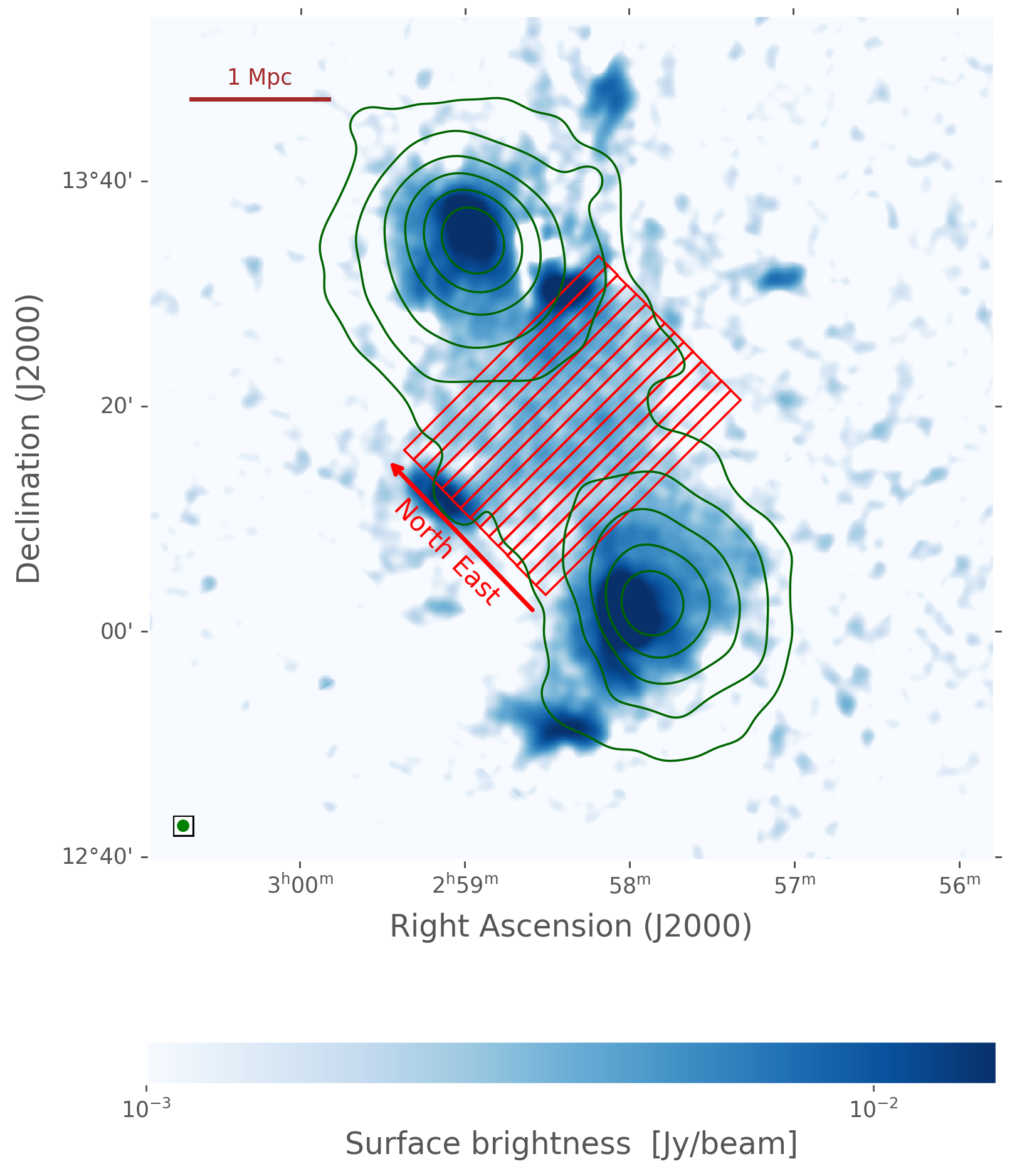}  
  \centering
  \includegraphics[width=0.48\textwidth]{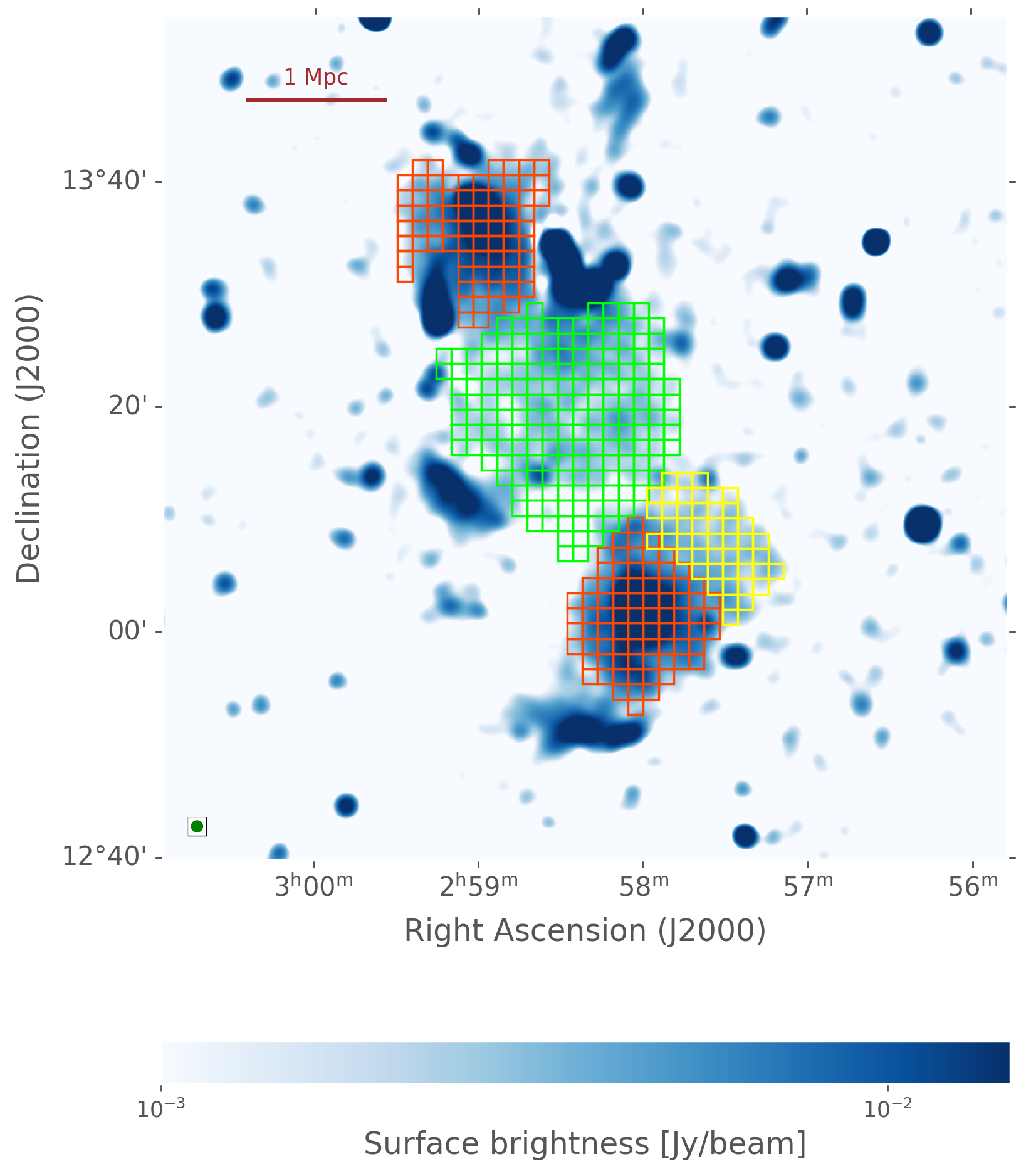} 
\caption{Radio maps of A399-401 where most compact sources are removed. \textit{Left panel}: Radio image in blue: \textit{uv}-subtract image described in the text. The square-root color scale
extends from $1\sigma$ to $25\sigma$. Green contour lines are from X-ray data from XMM Newton, which are smoothed with a Gaussian kernel. The red slices ($1100\arcsec\times75\arcsec$) point northeast. \textit{Right panel}: Radio image in blue: R02 filtered image described in the text. The square-root color scale
of the images extends from $1\sigma$ to $25\sigma$. It has a grid overlay on the two clusters and on the radio bridge on top of the radio image with the R02 filter. The orange grids cover the halos, the green grid covers the radio bridge, and the yellow grid covers the northwest radio halo extension from A399. The cell size in the grid is $80\arcsec\times 80\arcsec$. The scale for 1\,Mpc at $z=0.072$ is given in both images in the upper left corner.}
\label{fig:bridges}
\end{figure*}

\begin{figure*}[ht]
  \centering
  \includegraphics[width=0.48\textwidth]{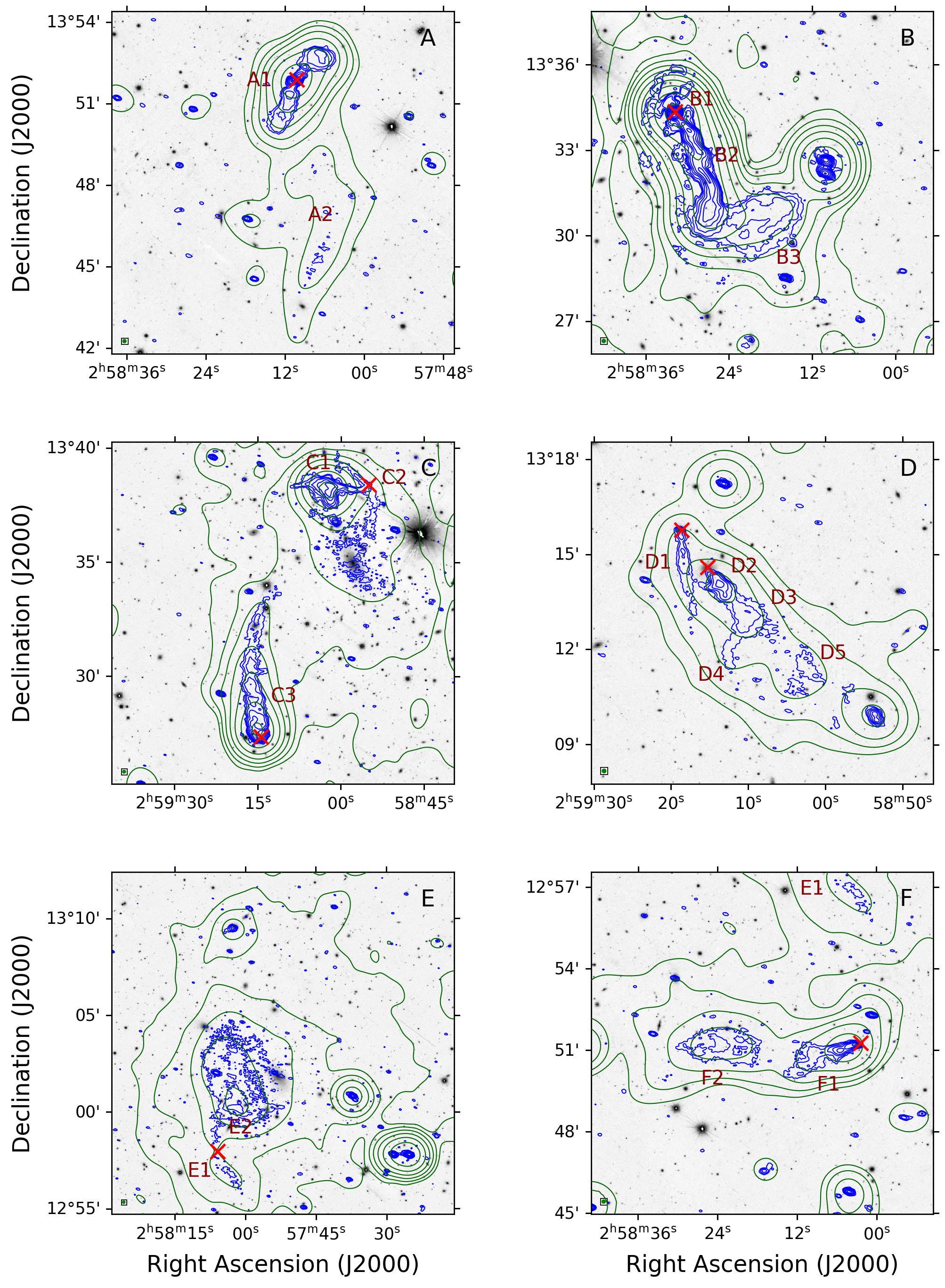}  
  \centering
  \includegraphics[width=0.48\textwidth]{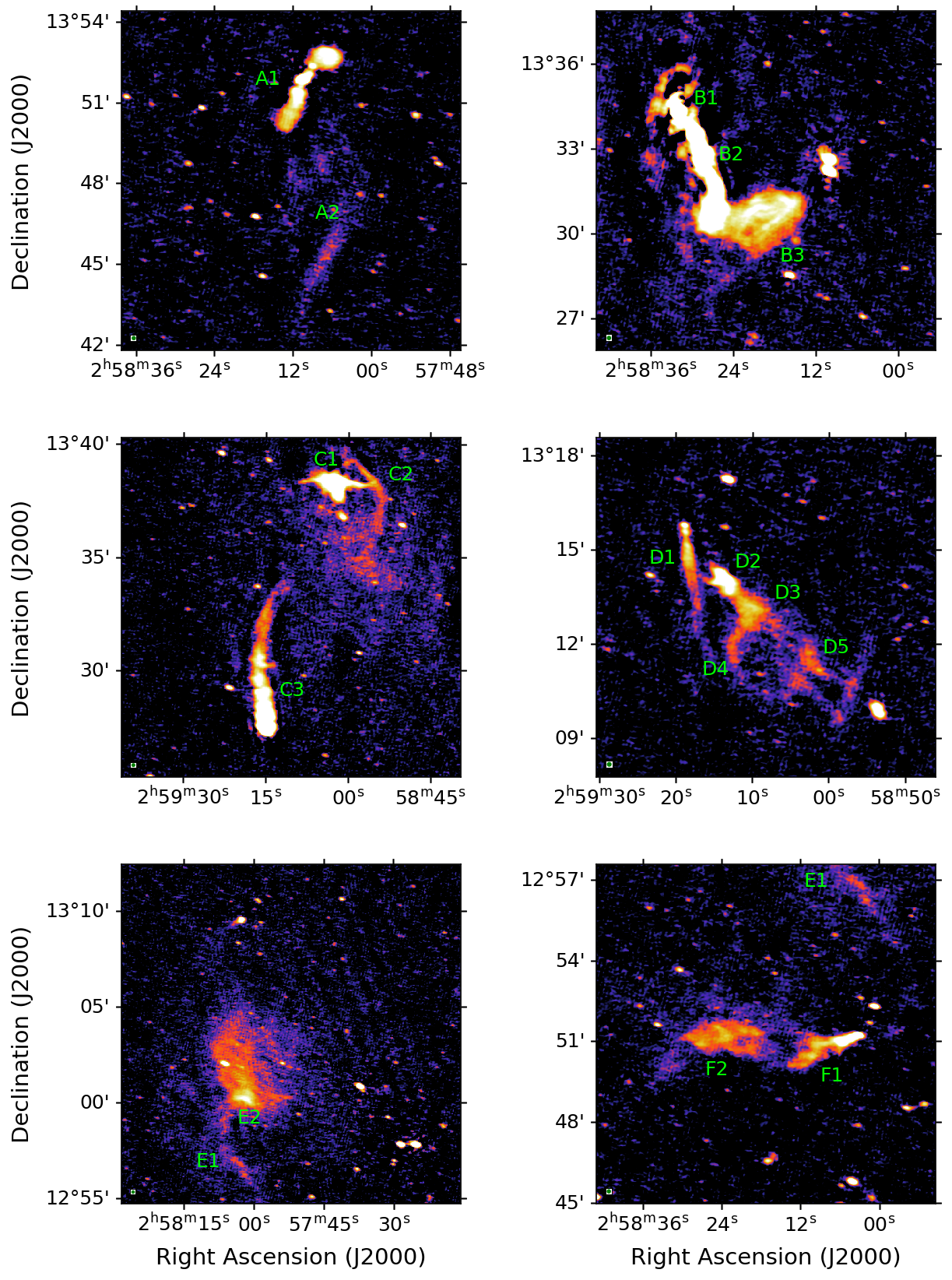}  
\caption{Zoomed images of the regions in Figure \ref{fig:6arc}. These are the same sources as in Fig. S4 from \cite{govoni2019} and ordered in the same way to facilitate comparison. \textit{Left two panels}: Blue contours at a resolution of $5.9\arcsec\times 10.5\arcsec$ and with $\sigma=79$ $\mu$Jy beam$^{-1}$ , and 
    green contours at a resolution of $72.8\arcsec\times75.9\arcsec$ and $\sigma=809$ $\mu$Jy beam$^{-1}$ , both at 144 MHz. Red crosses are elliptical host galaxies from the radio sources, and we label the radio components discussed in the text in brown. The background grayscale images are optical sources from Pan-STARRS DR1 \citep{panstarrs}. \textit{Right two panels}: Same regions as in the left panel at $5.9\arcsec\times 10.5\arcsec$ with a surface brightness color plot. The square-root color scale
of the images extends $1\sigma$ to $25\sigma$ to better highlight the plasma, and the labels are the same as in the left panel in green.}\label{fig:subimages}
\end{figure*}

Figures \ref{fig:6arc} and \ref{fig:lowres} present the LOFAR observation of A399-401 at three different resolutions. In Figure \ref{fig:bridges} we show the \textit{uv}-subtract and R02 filtered images, with grids and slices that are explained below. The radio map from our highest-resolution map is more than four times deeper than the 10$\arcsec$ radio map from \cite{govoni2019} ($\sigma=79$ $\mu$Jy beam$^{-1}$ versus $\sigma=300$ $\mu$Jy beam$^{-1}$). 
We also show zoomed images in Figure \ref{fig:subimages}, similar to Figures S3 and S4 in \cite{govoni2019}, and use labels in Figure \ref{fig:6arc} and \ref{fig:subimages} to mark a number of particularly interesting regions that are referred to throughout this paper.

\subsection{Diffuse emission}

In Figure \ref{fig:6arc} we clearly detect two radio halos belonging to A401 in the north and A399 in the south. At the lower resolutions in Figure \ref{fig:lowres}, we also directly observe the diffuse emission from the radio bridge. Below we discuss the morphology of the radio bridge and halos and measure the flux densities and radio powers.

\subsubsection{Morphology}\label{section:radiomaps}

\begin{figure}[ht]
  \centering
  \includegraphics[width=0.48\textwidth]{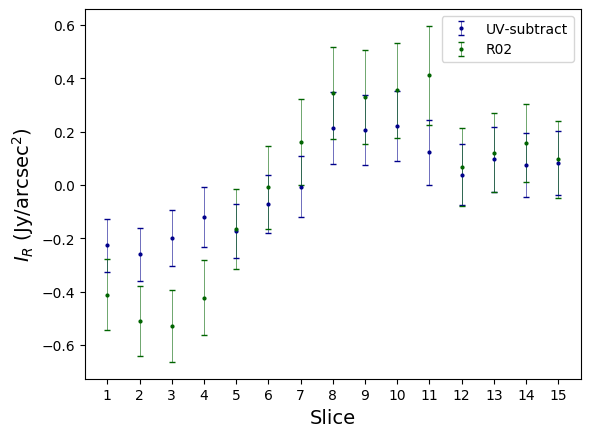}
\caption{Mean radio surface brightness from the slices from the left panel in Figure \ref{fig:bridges}, manually masked for bright AGNs, for the R02 filter and \textit{uv}-subtracted radio data. The slice numbers correspond to the order of the slices in the direction from the arrow. The R02 filter removes more from the diffuse emission than the \textit{uv}-subtract. Error bars include the statistic and systematic uncertainties.}
\label{fig:slicesgraph}
\end{figure}

In the two panels from Figure \ref{fig:lowres}, we observe a prominent brightness depression west of the bridge close to A399. We can detect this depression also from slices 1 to 5 in the radio surface brightness profile in Figure \ref{fig:slicesgraph}, where we slice through the radio bridge images in Figure \ref{fig:bridges} in the northeast direction for both the R02 filter and \textit{uv}-subtract radio maps. We identify various compact radio sources in the bridge area, but with the exception of the sources in region D from Figure \ref{fig:subimages}, we do not detect any indication of a morphological relation between them and the radio bridge.

In Figure \ref{fig:subimages} we highlight the radio halo from A401 in region C. The largest linear size (LLS) for this radio halo, measured within the $3\sigma$ contour, is 1.6\,Mpc. In the region covered by the radio halo, we detect a morphologically complex bright source (C1) that has no direct optical counterpart. This structure has a straight feature with a full LLS of 300\,kpc on its western and eastern side, but on its western side it, seems to connect to a bent jet-like structure (C2) that might be associated with the elliptical galaxy 2MASX\,J02585500+1338243 at $z=0.079$ \citep{hill1993}. 
In region E from Figure \ref{fig:subimages}, we see the radio halo from A399. From a $3\sigma$ isophote, we find an LLS of 1.3\,Mpc. In the southern area, attached to the halo, we detect a diffuse bent structure with two components with an LLS of \textasciitilde\,230\,kpc and \textasciitilde\,150\,kpc (E1). In the middle of these components, we can notice a small dip in the emission where we find the elliptical galaxy 2MASX\,J02580602+1257559 at $z=0.075$ \citep{hill1993}. This dip, at the location of the optically detected galaxy, suggests that this is a switched-off radio galaxy. The radio halo also has a brightness enhancement without an optical counterpart (E2) with an LLS of \textasciitilde\,100\,kpc. The lower resolutions in Figure \ref{fig:lowres} show that the radio halo extends to the northwest, which we treated as a separate component for the further analysis and in the discussion in Sections \ref{section:ptp} and \ref{section:discussion}.

\subsubsection{Flux densities and radio powers}

To calculate the integrated flux densities for the halos, we fit the radio halos with the halo-flux density calculator (\texttt{Halo-FDCA}) \citep{boxelaar2021}. This software applies a Markov chain Monte Carlo (MCMC) method to fit an exponential surface brightness profile, $$I(r)=I_{0}e^{G(\mathbf{r})},$$
where $I_{0}$ is the central radio surface brightness, and $G(\mathbf{r})$ is a quantity related to the morphology of the halo as a function of the two-dimensional distance from the center ($\mathbf{r}$). The radio power is calculated by $$P_{\text{144 MHz}}=\frac{4\pi D_{L}^{2}}{(1+z)^{1+\alpha}}S_{\text{144 MHz}},$$
where $D_{L}$ is the luminosity distance and $S_{\text{144 MHz}}$ is the integrated total flux for a spectral index $\alpha$. 
For A399 and A401, we find that a circular model with $G(\mathbf{r})=-\frac{\mathbf{r}}{r_{e}}$ works well, where $r_{e}$ is the \textit{e}-folding distance to the halo center. With this fit, we can estimate the flux density and radio power with corresponding uncertainties. Following \cite{murgia2009}, we decided to integrate up to $3r_{e}$. For the radio halo from A399, we obtain $r_{e}=208\pm 6$\,kpc and find $S_{\text{144 MHz}}=0.98 \pm 0.10$\,Jy and $P_{\text{144 MHz}}=0.99 \pm 0.11 \times 10^{25}$\,W Hz$^{-1}$ by using the best estimated spectral index \citep[$\alpha=1.75\pm 0.14$;][]{nunhokee2021}. For the radio halo in A401, we obtain $r_{e}=208\pm 7$\,kpc and find $S_{\text{144 MHz}}=0.77 \pm 0.08$\,Jy and $P_{\text{144 MHz}}=0.99 \pm 0.03 \times 10^{25}$\,W Hz$^{-1}$ with the best estimated spectral index \citep[$\alpha=1.63 \pm 0.07$;][]{govoni2019}. The output from \texttt{Halo-FDCA} is shown in Appendix \ref{appendix:halofdcaresults}.
 \texttt{Halo-FDCA} is created for radio halos. For the radio bridge, we therefore created a manual region that we associated with the bridge (covering a similar area as the green boxes in the right panel in Figure \ref{fig:bridges}) and integrated over this area in the \textit{uv}-subtract image. We obtain $S_{\text{144 Mhz}}=0.55 \pm 0.06$\,Jy, and by adopting the current lowest estimated spectral index for the bridge \citep[$\alpha=1.5$;][]{nunhokee2021}, we find $P_{\text{144 MHz}}=0.75 \pm 0.08 \times 10^{25}$\,W Hz$^{-1}$ as the upper limit. 
All values are listed together in Table \ref{table:properties}. The uncertainties include systematic, subtraction, and statistical errors.
The systematic error takes into account the uncertainty of the flux scale calibration, the subtraction error takes into account the uncertainty from remaining residual emission from discrete sources after subtraction, and the statistical error takes takes into account the sensitivity of the image \citep[see also Section 5 from ][]{botteon2022}.

\cite{murgia2010} also used a circular exponential fit to calculate the flux density for the radio halo from A399. They found  $r_{e}=186 \pm 16$\,kpc at 1.4 GHz with VLA data, which is lower than but still consistent within the error bars with our value.
\cite{govoni2019} used a different radio map, with a lower resolution and sensitivity than our map ($10\arcsec$ and $\sigma=300$ $\mu$Jy beam$^{-1}$ respectively), to measure the flux densities from the radio halos over a $5\sigma$ isophote.
Despite these different methods, our values for the radio halos are consistent within the error bars with those from \cite{govoni2019} ($S_{\text{140 MHz}}=826\pm 126.5$ mJy for A401; $S_{\text{140 MHz}}=807\pm 124.7$ mJy for A399). \cite{govoni2019} measured $S_{\text{140 MHz}}=822\pm 147$ mJy over 3.9 Mpc$^{2}$ for the radio bridge. This area is a significantly larger than the 2.7 Mpc$^{2}$ that we find for the bridge. Our bridge area is more conservatively chosen than \cite{govoni2019} because they extrapolated the average surface brightness from cluster core to cluster core for the masked regions (radio halos and sources), while we excluded the radio halos from the bridge area entirely. For the average surface brightness, we both find \textasciitilde 0.38 $\mu\text{Jy arcsec}^{-2}$, which means that our results (independent of the area we constrain for the bridge) are consistent with each other.

\begin{table*}
 \caption{Measured physical properties at 144 Mhz with spectral indices from \cite{govoni2019} and \cite{nunhokee2021}.}
 \label{table:properties}
\begin{center}
\begin{tabular}{l l l l l}
 \hline
& $r_{e}$ [kpc] & $S_{\text{144 Mhz}}$ [Jy] & $P_{\text{144 MHz}}$ [W/Hz] & $\alpha$ \\
 \hline\hline
\bf{A399} & $208\pm 6$ & $0.98 \pm 0.10$ & $1.26 \pm 0.13 \times 10^{25}$ & $1.75 \pm 0.14$ \\ 
\bf{A401} & $241 \pm 7$ & $0.77 \pm 0.08$ & $0.99 \pm 0.11 \times 10^{25}$ & $1.63 \pm 0.07$ \\ 
\bf{Bridge} & N.A. & $0.55 \pm 0.06$ & $0.75 \pm 0.08 \times 10^{25}$ & $>1.5$ \\ 
\hline
\end{tabular}
\end{center}
\end{table*}

\subsection{AGNs}\label{sec:agn}

In our radio images, we detected several interesting radio components that are associated with galaxies in or near A399-401. These bright radio sources make up a large part of all the radio emission in A399-401. We therefore discuss the radio components in our radio maps in detail.
Most of the radio emission is likely associated with AGNs, as already discussed by \cite{govoni2019}. Remnant plasma from AGN are also a possible ingredient to explain the origin of the emission from the radio bridge and radio halos (this is further discussed in Section \ref{section:discussion}).

In region A from Figure \ref{fig:subimages}, we detect the radio galaxy (A1) corresponding to 2MASX\,J02581042+1351526, which was associated by \cite{harris1980} as a probable member of A401 based on its distance to the cluster center (\textasciitilde 1.5 Mpc) and brightness. This object has two lobes with a full LLS of \textasciitilde\,300\,kpc, and it is connected to diffuse emission pointed toward the A399-401 system (A2). Region B in Figure \ref{fig:subimages} has a bright active radio core (B1) corresponding to the elliptical galaxy 87GB\,025547.6+132220 at $z=0.084$ \citep{huchra2012}. Whereas there was a gap between the long tail (B2) and the core (B1) in the images from \cite{govoni2019}, we detect it as one connected structure. 
The radio emission from B2 has an LLS of \textasciitilde\,320\,kpc. Connected at the end of this emission, we detect remnant plasma stretched to the west over a similar LLS of \textasciitilde\,270\,kpc (B3). We observe signs of plasma extending southward into the radio bridge filament at the edges
of B2 and B3.
To the southeast of the radio halo from A401, we detect a radio galaxy (C3) in region C from Figure \ref{fig:subimages} that we associate with 2MASX\,J02591487+1327117 at $z=0.078$ \citep{hill1993}. Component C3 has a \textasciitilde\,550\,kpc long prominent tail extending toward the radio halo. 

In region D in Figure \ref{fig:subimages}, we detect two sources next to each other: 2MASX\,J02591878+1315467, located  northeast at $z=0.073$, and southwest from this source, we find 2MASX\,J02591535+1314347 at $z=0.078$ \citep{hill1993}. The first is an elliptical galaxy associated with a tail with an LLS of \textasciitilde\,220\,kpc (D1), and the second is an elliptical galaxy associated with the bright radio core to the southwest (D2) with an LLS of \textasciitilde 90\,kpc. From the core, a fainter component extends southwest (D3) with an LLS of \textasciitilde\,80\,kpc and a long bent structure pointed to the south (D4). Following the same direction to the southwest, the diffuse emission again becomes brighter (D5), and in the lower-resolution images in Figure \ref{fig:lowres}, this area seems to be connected to the bridge. No optical galaxy is associated with this emission. 
In region F in Figure \ref{fig:subimages}, immediately below the radio halo from A399, we detect two regions with brighter plasma that lie next to each other. While the \textasciitilde\,250\,kpc long plasma on the west side (F1) can be associated with the optical elliptical galaxy 2MASX\,J02580300+1251138, which is located at $z=0.075$ \citep{hill1993}, we do not find an obvious optical counterpart for F2, which has about the same length. Component F1 corresponds to a tailed source, and we suspect that F2 might be the extension of F1, given its morphology (see the discussion in Section \ref{section:radiohalos}).

\section{Thermal and nonthermal scaling relations}\label{sec:thermalnonthermal}

It has been shown observationally \citep[e.g.,][]{cassano2013, vanweeren2021} that the thermal emission from the ICM and nonthermal radio emission are related by the following scaling relation: 
$$P_{\nu} \propto M^{\gamma},$$
which is the relation of the radio power integrated over the entire cluster $P_{\nu}$ at a wavelength $\nu$ and the cluster mass $M$ derived from X-ray or SZ measurements with slope $\gamma$. It was suggested that nonthermal emission is powered by the dissipation of gravitational energy \citep[e.g.][]{cassano2006}. 

Instead of using a statistical population of clusters to determine the thermal and nonthermal relation in the ICM by means of integrated quantities ($P$ and $M$), we can also use spatially resolved quantities on single objects by inspecting the following scaling relation:
$$I_{R}\propto I_{X}^{a},$$
where $I_{R}$ is the radio surface brightness, and $I_{X}$ is the X-ray surface brightness with slope $a$. This relation has been derived for many radio halos with a point-to-point analysis \citep{govoni2001, feretti2001, giacintucci2005, rajpurohit2018, botteon2020b, rajpurohit2021, ignesti2020, biava2021, duchesne2021, bonafede2021}. A point-to-point analysis has also been performed for the radio bridge of A1758 \citep{botteon2020}. 
A strong correlation reflects the physics of the interplay between the thermal and nonthermal components \citep[e.g.,][]{brunetti2004, brunetti2014}, where the particle density and magnetic field strength (traced by the radio surface brightness) decline faster than the thermal gas density (traced by the X-ray surface brightness) if $a>1,$ or vice versa, if $a<1$.
We study these scaling relations for A399-401 below.

\subsection{Mass-power relation}

\begin{figure}[t]
  \centering
  \includegraphics[width=0.48\textwidth]{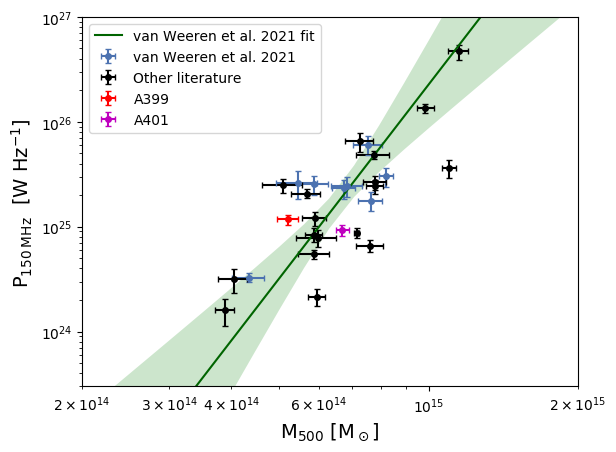}  
\caption{Mass and radio power relation. The fitted regression line with a $3\sigma$ confidence interval in the shaded area and the data points in cyan come from the sample in \cite{vanweeren2021}. The literature data points in black are taken, similar to \cite{vanweeren2021}, from previous LOFAR and \textit{GMRT} studies, and a correlation line is fit using the bivariate correlated errors and intrinsic scatter (\texttt{BCES}) orthogonal regression algorithm \citep{bces1996}. We left out the radio halo candidates. The radio halos A399 and A401 are added. Error bars include statistical and systematic uncertainties.}\label{fig:masspower}
\end{figure}

To determine where the radio halos from A399 and A401 are located in the cluster mass radio power diagram, we used the most recent relation at a close frequency from \cite{vanweeren2021}. 
Following \cite{cassano2013}, they derived the following scaling relation:
$$P_{\text{150 MHz}}\approx M_{500}^{6.13 \pm 1.11}$$
for a sample of clusters in a rest-frame of $P_{\text{150 MHz}}$ , and where $M_{500}$ is the cluster mass within a radius whithin which the average density is equal to 500 times the critical density of the Universe, taken from the PSZ2 \textit{Planck} catalog \citep{planck2016b}.
We included A399 and A401 in the figure from \cite{vanweeren2021} (see Figure \ref{fig:masspower}). The radio powers from Table \ref{table:properties} were scaled to 150 MHz using the spectral indices from \cite{govoni2019}. To be consistent with \cite{vanweeren2021}, we used the masses from \cite{planck2016b} ($5.24^{+0.29}_{-0.23} \times 10^{14}$ M$_{\odot}$ for A399; $6.75^{+0.22}_{-0.17} \times 10^{14}$ M$_{\odot}$ for A401). The radio halos are located close to the best-fit relation, implying that their radio powers are similar to those of other clusters with similar masses. 

 Table \ref{table:properties} indicates that the radio power of the bridge is comparable to that of both of the radio halos, while we know from \cite{hincks2022} that the mass of the bridge is only roughly 8\% of the total mass of A399-401. This means that the bridge does not fit into the same mass-power relation as the radio halos, which is no surprise if bridges and radio halos have different origins through different physical mechanisms. 
More observations of radio bridges are needed to infer whether a different mass-power relation exists.

\subsection{Point-to-point analysis}\label{section:ptp}

The spatial correlation between the radio and X-ray emission reflects the strength of the connection between the thermal and nonthermal components in the ICM. Therefore, a point-to-point analysis can be used as a tool for comparing the radio and X-ray emission and for obtaining information about the mechanisms of acceleration and transport of relativistic particles, and the amplification of magnetic fields in radio halos and bridges. This helps us to understand the origin of the radio bridge and halos better.

Similarly to the radio power and mass relation, we derived the radio and X-ray relation in log-log space,
$$\log\left(I_{R}\right) = a\log\left(I_{X}\right)+b,$$
where we refer to $a$ as the slope.
To quantify the correlation, we derived the Spearman correlation coefficient ($r_{\text{s}}$). 
We used the \textit{uv}-subtract and the R02 filtered map, which each have their advantages and disadvantages, as we explained in Section \ref{section:compact}. We generated a grid that covers the radio bridge and radio halos separately (see the right panel in Figure \ref{fig:bridges}). There is no clear boundary between the radio halos and the radio bridge, therefore we used a $5\sigma$ radio contour around the radio halos to define the border between the radio bridge and halo areas. Furthermore, the northwestern radio halo extension from A399 and the core from A399 each have their own grids (yellow and orange, respectively) because we show below that this will help to explain the origin of the radio halo from A399. We chose a grid cell size of $80\arcsec\times80\arcsec$. This size is slightly larger than the beam size for the radio and X-ray map and therefore prevents a correlation between contiguous cells in the grid. Larger cell sizes improve the signal-to-noise ratio but reduce the number of data points for a linear regression, especially for radio halos where there is less area to cover. 
We calculated the average surface brightness and errors for every cell, including the statistical and systematic uncertainties for the radio and X-ray emission. From the X-ray surface brightness, we subtracted the sky background contribution, which we derived to be $1.27\cdot 10^{-7}$ count/s/arcsec$^{2}$ in the 0.5-2.0 keV band.
This contributes up to \textasciitilde 40\% of the emission at the more diffuse edge of the bridge region from A399-401.
We adopted a radio surface brightness threshold of $2\sigma$ and removed cells covering areas that are not related to the radio halos or bridge, such as objects associated with AGNs, which are only partially removed in the source subtraction.
The $2\sigma$ threshold is needed to prevent any effect from unreliable flux densities and noise at a low signal-to-noise ratio.
Adopting a higher threshold can effectively flatten the slope values. To reduce this effect, we followed \cite{botteon2020b} and used \texttt{LIRA}, which is a Bayesian regression method that allows fitting data points that have a (2$\sigma$) threshold in the y-variable \citep{sereno2016}.\footnote{\url{https://cran.r-project.org/web/packages/lira/}}
With this regression method, we obtain a mean and error that reflect the errors on the radio and X-ray measurements as well.

\begin{figure*}[ht]
  \centering
  \includegraphics[width=0.48\textwidth]{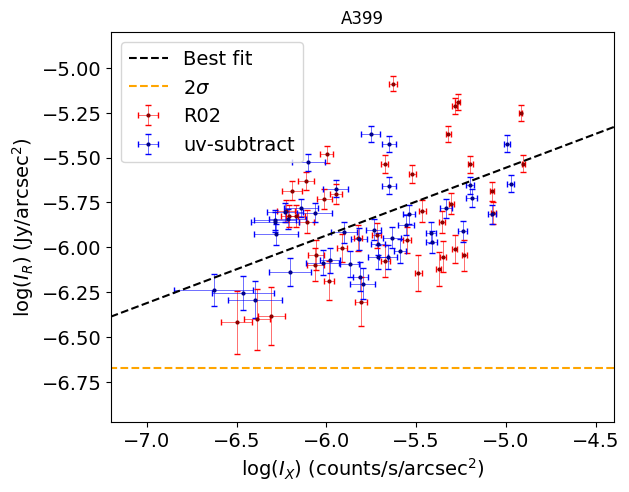}  
  \centering
  \includegraphics[width=0.48\textwidth]{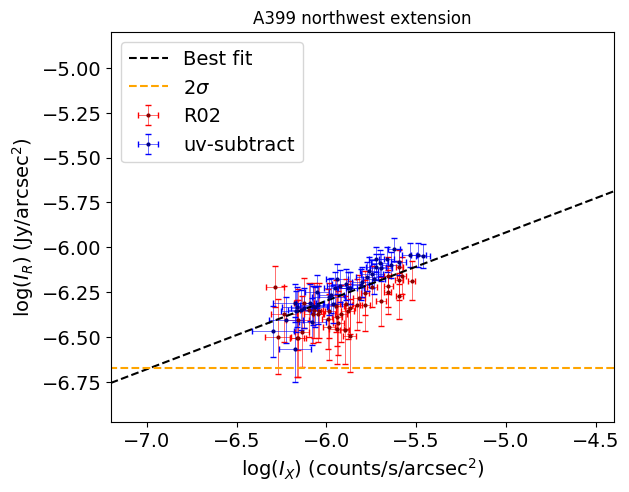}  
  \centering
  \includegraphics[width=0.48\textwidth]{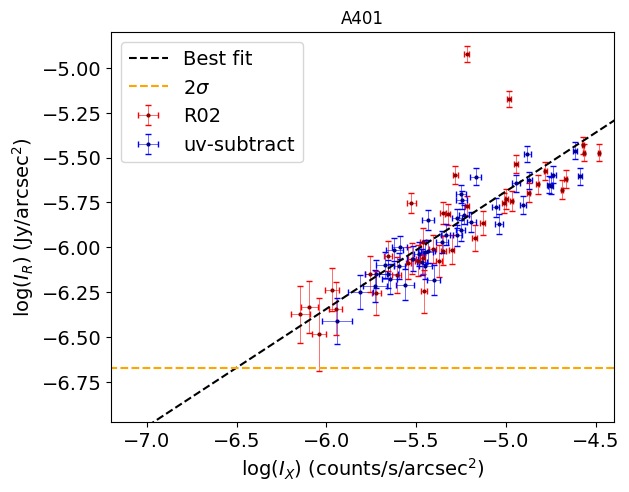}
  \centering
  \includegraphics[width=0.48\textwidth]{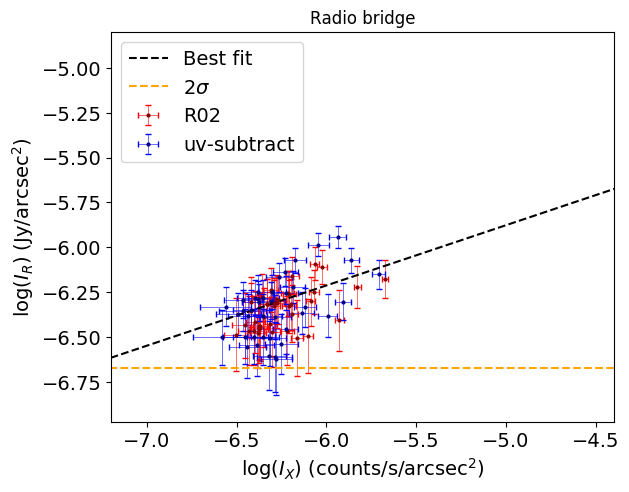}  
\caption{Radio and X-ray surface brightness correlation plots for every cell from the grid used for the point-to-point analysis in the right panel of Figure \ref{fig:bridges} for the R02 filter and \textit{uv}-subtract maps. This is just one grid from the MC grid generation with 40 randomly drawn points to improve the readability of the plot. \textit{Top left panel}: Radio halo of A399. \textit{Top right panel}: Radio halo extension of A399. \textit{Lower left panel}: Radio halo of A401. \textit{Lower right panel}: Radio bridge.
Error bars include the statistic and systematic uncertainties. The best linear fit with \texttt{LIRA} is given with a dashed black line. Only radio brightness values higher than $2\sigma$ are included.}
\label{fig:correlations}
\end{figure*}

To reduce the choice sensitivity of the grid, we generated grids in a Monte Carlo (MC) fashion. This is similar to what is described in \cite{ignesti2022}. In our approach, we made multiple grid realisations with small random offsets of $40\arcsec$ at most (half of the grid resolution) around the starting and ending points of the grid. In this way, we generated 100 different grid layouts for the R02 filtered and \textit{uv}-subtract radio maps. An example of a grid layout is shown in Figure \ref{fig:bridges} (right panel). 
Finally, we calculated the final values and errors for the slopes and the Spearman correlation coefficients with the bootstrapping method similar to \cite{ignesti2022}, such that we propagated the errors from individual fits to the final values. All results are presented in Table \ref{table:corr}. The correlation plots are shown for one particular grid choice in Figure \ref{fig:correlations}. 

\begin{table*}
 \caption{Correlation coefficients and slopes for $I_{R}-I_{X}^{a}$.}
 \label{table:corr}
\begin{center}
\begin{tabular}{l c c c c c }
 \hline
  & $r_{\text{s}}$ & $a$ \\ [0.5ex] 
 \hline\hline
\bf{A399} & $0.60\pm0.20$ & $0.33 \pm 0.11$\\ 
\bf{A399} core & $0.47\pm0.15$ & $0.32 \pm 0.07$\\ 
\bf{A399} HE & $0.71 \pm 0.15$ & $0.35 \pm 0.12$\\
\bf{A401} & $0.91\pm 0.04$ & $0.63 \pm 0.06$\\
\bf{Bridge} & $0.41\pm0.14$ & $0.27\pm 0.07$\\ 
\end{tabular}
\end{center}
\end{table*}

All correlations are positive and all slopes are sublinear ($a<1$) in Table \ref{table:corr}. 
The radio halo extension from A399 has a much steeper slope and a stronger correlation between the radio and X-ray emission than the core from A399. In Section \ref{section:radiohalos} we provide more detail about this.

\section{Discussion}\label{section:discussion}

Because radio bridges are a very recent discovery, the origin of the radio emission from these structures remains an open question. 
The Mpc scale size of the A399-401 radio bridge and the maximum sub-Mpc distances that relativistic particles can travel due to age constraints require an in situ mechanism as the driver of the emission of synchrotron radiation throughout the bridge region \citep{brunetti2014}. 
An important ingredient for these models is the presence of fossil cosmic-ray electrons \citep[e.g.,][]{brunetti2001, brunetti2011, petrosian2001, pinzke2017}. These seed relativistic electrons in the energy range 100-500 MeV have a long lifetime in the ICM and can be injected by past shock activity, AGNs, galactic winds, or via the decay of charged pions from proton-proton collisions \citep[e.g.,][]{brunetti2014}. During merger events, fossil cosmic rays can be reaccelerated via various Fermi-I or II mechanisms and/or be reenergized by adiabatic compression \citep[see][for a review]{brunetti2014}. Evidence for the revival of AGN fossil radio plasma, for example, so-called gently reenergized tails (GReEt) and radio phoenices, has been observed in a number of clusters \citep[e.g.,][]{gasperin2017,vanweeren2017,mandal2020}.

As the radio bridge is connecting the radio halos from A399 and A401, it is important to understand the origin of these radio halos as well. The fact that we observe radio halos in the center of these clusters also means that that they are also undergoing their own mergers \citep{buote2001, cassano2010}.

\subsection{Origin of the radio bridge}\label{sec:bridge}

Although \cite{govoni2019} initially suggested a model in which weak shocks reaccelerate particles via a Fermi-I type mechanism, recent studies favor a model with Fermi-II reacceleration through turbulence to explain the origin of the A399-401 radio bridge \citep{brunetti2020}.
With the point-to-point analysis, we find a trend between the radio and X-ray surface brightnesses, similar to \cite{botteon2020}. We also find a source for fossil plasma, as we detect evidence for AGN injection of relativistic particles into the radio bridge region. In particular, we observe radio brightness enhancements around AGN jets that are likely places where plasma is being injected in the radio bridge. This is best visible at the southern tip of components B2 and B3 (see Figure \ref{fig:subimages}). Moreover, the plasma from A2, likely coming from an AGN (see Section \ref{sec:agn}), is pointed toward the bridge. These examples make a compelling case for the scenario in which in the past, these and other AGNs have dumped fossil plasma, which now functions as the source of primary seed electrons ready to be reaccelerated through turbulence. At the same time, the fossil plasma can scatter the radio surface brightness distribution, which in return can reduce the spatial correlation between the radio and X-ray emission.
With the fossil plasma we also have an important ingredient for an in situ turbulent reacceleration model. In addition, we do not observe filamentary structures or shock surfaces in the bridge region, which is another indication that the emission is volume filling, and turbulence instead of shock models is currently the best explanation. Fermi-II reacceleration therefore remains the best candidate to explain the origin of the radio bridge.
We suggest that in follow-up work, a high-resolution study of the spectral index across the bridge would be interesting to better understand the fate of the relativistic electrons that are injected into thermal gas and the role of the reacceleration processes.

An observed X-ray temperature break by \cite{akamatsu2017} in the region that we labeled D in Figure \ref{fig:subimages} is suggested to be a sign of an equatorial shock from the A399-401 merger axis \citep{ha2018} or a milder adiabatic compression between the clusters \citep{gu2019}. Figure \ref{fig:subimages} shows that the emission from D5 is stretched in the same direction as the jet from D3, which is associated with an AGN in 2MASX\,J02591535+1314347. Therefore, we propose that the emission from D5 is a remnant tail from the AGN, which is reenergized by the equatorial shock or adiabatic compression. This is an alternative to the candidate radio relic classification by \cite{govoni2019} or the switched-off radio galaxy explanation from \cite{nunhokee2021}. We further observe in Figure \ref{fig:lowres} that the emission from D5 is directly connected with the bridge. This shows that the morphology of the radio bridge could be directly affected by this shock or compression, which can also change the spatial correlation between the radio and X-ray emission. 

To determine how the values from the point-to-point analysis for A399-401 compare with other radio bridges, we considered the point-to-point analysis that was performed for the radio bridge in A1758. We used the data behind Figure 4 from \cite{botteon2020} and derived $a=0.25\pm 0.08$ and $r_{\text{s}}=0.52\pm0.22$. These values are similar to what we find for the radio bridge in A399-401 (see Table \ref{table:corr}). Although we used a finer grid resolution (\textasciitilde 115 kpc versus \textasciitilde 185 kpc) than \cite{botteon2020}  for A399-401, this might indicate that the two radio bridges are produced by similar processes. When we compare the slopes from these two radio bridges with mini and giant radio halos in the literature or with those from A399 and A401, we conclude that slopes from radio bridges are overall flatter \citep{govoni2001, feretti2001, giacintucci2005, rajpurohit2018, rajpurohit2021, botteon2018, botteon2020, ignesti2020, biava2021, bonafede2021, bonafede2022}. This might be an indication that the physical connections between the ICM and radio bridges and the ICM and radio halos are different from each other. More radio bridges in premerging clusters need to be studied to conclude whether these correlation and slope values are typical for radio bridges between premerging clusters, and how this relates to the underlying physical processes.

\subsection{Radio halos}\label{section:radiohalos}

With our point-to-point analysis, we find a sublinear slope and a remarkable strong correlation between the radio and X-ray emission from A401. Similar to the radio bridge, we also detect AGNs that inject the radio halo environment with plasma. First of all, Figure \ref{fig:subimages} shows that the AGN tail from C3 is directed toward the bright enhanced part in the radio halo. Second, the AGN labeled C2 affects the radio halo environment from the edges from its southern lobe. The combination of a strong radio and X-ray correlation with the observed AGN activity suggests a scenario in which turbulent Fermi-II reacceleration of fossil plasma injected by AGNs in the cluster causes most of the radio emission from this radio halo \citep{brunetti2008, brunetti2014, zuhone2015}. This is further supported by the steep spectrum ($\alpha=1.75 \pm 0.14$) measured by \cite{nunhokee2021}, which can be best explained by a turbulent reacceleration mechanism in moderately disturbed systems \citep{brunetti2008}.
 
We find a weaker correlation between the radio and X-ray emission for the radio halo from A399 (with and without the northwest extension) than for the radio halo from A401 with the point-to-point analysis. The relation between the radio and X-ray emission components is likely affected by the cluster merger in A399 between a higher-mass system and a lower-mass system going from east to west, as proposed by \cite{sakelliou2004} and simulated by \cite{takizawa1999}. Evidence for this merger comes from an X-ray edge at the southeast side of the cluster core of A399 \citep{sakelliou2004}. The edge coincides with the region of enhanced radio brightness at the center of the radio halo, where the enhanced emission labeled E2 in Figure \ref{fig:subimages} might be the result of a weak shock from the merger event \citep{murgia2010}. The unrelaxed dynamical state of the cluster is reflected in the offset of the radio halo peak with respect to the X-ray peak. This is also in line with the cold front claimed by \cite{botteon2018b}. Other clusters that are in a complex merging stage show similar weaker correlations between the radio and X-ray emission from a point-to-point analysis \citep{shimwell2014, duchesne2021}.
In contrast, we find a steeper slope and a strong radio and X-ray correlation in the northwest extension (labeled HE in the lower right panel in Figure \ref{fig:lowres}) from the radio halo. Together with steep-spectrum from \cite{nunhokee2021}, this makes a case for turbulent Fermi-II reacceleration of cosmic rays in A399, which is directed from the merger axis toward the northwest from the radio halo \citep{brunetti2008}, and where a recent merger between a higher- and lower-mass system scatters the radio and X-ray relation around the core.
The two radio tails from E1 (Figure \ref{fig:subimages}) seem to be coming from a currently switched-off AGN, as the optical source is situated in the brightness dip between these jets. Its northern jet is directly connected with the radio halo and might be a source of seed particles that are needed for the turbulent reacceleration in the radio halo. Instead, it is also possible that fossil plasma in the jets is reenergized by the currently ongoing merger in A399. Farther south of this cluster, we detect emission labeled F2 (Figure \ref{fig:subimages}), which might be a reenergized fossil plasma (originating in but disconnected from the AGN tail from 2MASX\,J02580300+1251138) by the merger in A399. This is again an alternative explanation to the candidate radio relic classification from \cite{govoni2019}.

\section{Conclusions}\label{section:summary}

A399-401 is one of a few giant intracluster radio bridges that have been observed so far \citep[e.g.,][]{govoni2019, botteon2020}. We created new radio maps from A399-401 by using the improved recalibration method from \cite{vanweeren2021} and combining this with the wide-field facet imaging mode in \texttt{WSClean} version 3 on \textasciitilde\,40h LOFAR data from six different observations. Despite the high computational costs compared to the standard \texttt{DDF-Pipeline}, we find that this method works well for calibrating large diffuse structures where calibration artifacts around compact sources can be an issue in reconstructing the diffuse emission with the \texttt{DDF-Pipeline}. In the case of A399-401, we measure improvements of a factor \textasciitilde 1.6 in dynamic range for bright compact sources in our recalibrated radio map compared with the radio map produced with the \texttt{DDF-Pipeline} . In comparison with the previously studied radio map of A399-401 \citep{govoni2019}, we improved the resolution from $10\arcsec\times10\arcsec$ to $5.9\arcsec\times10.5\arcsec$ and the sensitivity from $300$ $\mu$Jy beam$^{-1}$ to $79$ $\mu$Jy beam$^{-1}$.

By analyzing the resulting images and using a point-to-point analysis to compare the radio and X-ray surface brightness changes across a region, we find the following:

\begin{itemize}
  \item We clearly detect the radio halos and the radio bridge in A399-401. We report for the first time a prominent brightness depression close to the radio halo from A399, starting west of the bridge. This shows that the radio bridge is not one straight elongated structure stretching from A399 to A401.
  \item We find a trend between the radio and X-ray emission for the radio bridge with a point-to-point analysis. We also detect radio surface brightness enhancements around bright AGN jets, which are an indication that fossil plasma has been left by past AGN activity. This might also scatter the radio surface brightness distribution and therefore weaken the correlation between the radio and X-ray emission in a point-to-point analysis. At the same time, this fossil plasma is necessary for in situ reacceleration. Together with the already constrained steep-spectrum \citep[$\alpha>1.5$;][]{nunhokee2021} from \cite{govoni2019}, these observations make a case for Fermi-II reacceleration to explain the origin of the radio bridge.
  \item We obtain similar results from the point-to-point analyses in the radio bridges in A1758 and A399-401. This suggests that these radio bridges might have similar origins.
  \item By applying the point-to-point analysis to the radio halo from A401, we find a strong correlation between the radio and X-ray emission. Together with signs of AGN activity in the radio halo and its steep spectrum \citep[$\alpha=1.63\pm 0.07$;][]{govoni2019}, we argue that it is likely that the emission from this halo originates in Fermi-II reacceleration.
  \item We see the effects of a recent merger in A399 in a weaker radio and X-ray correlation compared to what we find for A401. However, we find a strong correlation in the northwest radio halo extension. We therefore argue that this observation, together with the steep spectrum from the radio halo in A399 \citep[$\alpha=1.75\pm 0.14$;][]{nunhokee2021}, is in favor of a scenario in which Fermi-II reacceleration through turbulence is the main mechanism to explain the origin of the emission.
  \item We suspect that the two earlier classified radio relics by \cite{govoni2019} might be reenergized fossil plasmas from earlier AGN activity. This supports the importance of reacceleration and fossil plasma as drivers of the diffuse emission in A399-401.
\end{itemize}

Our work shows the power of refining the calibration and imaging of data from LOFAR to help us to study the diffuse emission between premerging clusters.
With our results, we can conclude that reacceleration through turbulence and current and past AGN activity are likely important ingredients to explain most of the radio emission in A399-401 and possibly other radio bridges as well. 

 \begin{acknowledgements} 
 
This publication is part of the project CORTEX (NWA.1160.18.316) of the research programme NWA-ORC which is (partly) financed by the Dutch Research Council (NWO). This work made use of the Dutch national e-infrastructure with the support of the SURF Cooperative using grant no. EINF-1287.
 
RJvW acknowledges support from the ERC Starting Grant ClusterWeb 804208. AB acknowledges support from the VIDI research programme with project number 639.042.729, which is financed by the Netherlands Organisation for Scientific Research (NWO), and from the ERC Starting Grant DRANOEL 714245. RC acknowledges support from INAF mainstream project ‘Galaxy Clusters Science with LOFAR’ 1.05.01.86.05.
 
This paper is based on data obtained with the International LOFAR Telescope (ILT). LOFAR \citep{haarlem2013} is the LOw Frequency ARray designed and constructed by ASTRON. It has observing, data processing, and data storage facilities in several countries, which are owned by various parties (each with their own funding sources) and are collectively operated by the ILT foundation under a joint scientific policy. The ILT resources have benefited from the following recent major funding sources: CNRS-INSU, Observatoire de Paris and Université d’Orléans, France; BMBF, MIWF-NRW, MPG, Germany; Science Foundation Ireland (SFI), Department of Business, Enterprise and Innovation (DBEI), Ireland; NWO, The Netherlands; The Science and Technology Facilities Council, UK; Ministry of Science and Higher Education, Poland; The IstitutoNazionale di Astrofisica (INAF), Italy. 
 
The Pan-STARRS1 Surveys (PS1) and the PS1 public science archive have been made possible through contributions by the Institute for Astronomy, the University of Hawaii, the Pan-STARRS Project Office, the Max-Planck Society and its participating institutes, the Max Planck Institute for Astronomy, Heidelberg and the Max Planck Institute for Extraterrestrial Physics, Garching, The Johns Hopkins University, Durham University, the University of Edinburgh, the Queen’s University Belfast, the Harvard-Smithsonian Center for Astrophysics, the Las Cumbres Observatory Global Telescope Network Incorporated, the National Central University of Taiwan, the Space Telescope Science Institute, the National Aeronautics and Space Administration under Grant No. NNX08AR22G issued through the Planetary Science Division of the NASA Science Mission Directorate, the National Science Foundation Grant No. AST-1238877, the University of Maryland, Eotvos Lorand University (ELTE), the Los Alamos National Laboratory, and the Gordon and Betty Moore Foundation.
 
 Based on observations obtained with XMM-Newton, an ESA science mission with instruments and contributions directly funded by ESA Member States and NASA.
 
 \end{acknowledgements}

\appendix

\section{Computing recalibration}\label{appendix:spider}

The images for A399-401 were produced with the recalibration method described in Section \ref{section:calibrationmethod}. We used 24 boxes within 0.6 degrees from the pointing center.
All the extractions and self-calibrations were done using processor nodes on Spider\footnote{\url{https://spiderdocs.readthedocs.io/}}, which is a high-throughput data-processing platform from SURF\footnote{\url{www.surf.nl}} and allows to run parallel jobs. 

\begin{figure}[ht]
  \centering
  \includegraphics[width=0.6\textwidth]{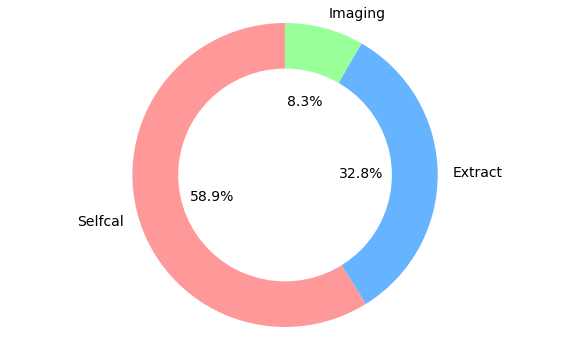}  
\caption{CPU core hours in percentage from the total amount for self-calibration, extraction, and imaging for making the image from A399-401. This is based on the calibration of A399-401 with 24 boxes within 0.6 degrees from the pointing center.}
\label{fig:corehours}
\end{figure}

The total number of CPU core hours (processor units multiplied by job hours) for the recalibration is 50336.
In Figure \ref{fig:corehours} we see that the self-calibration almost used 2/3th of the total CPU core hours to process the data, while the extraction used around 1/3th. The imaging is the smallest component in the recalibration. The self-calibration and extraction step for every individual box can only be run in a serial manner, while the boxes can run in parallel of each other. This means that in the optimal case, we can speed-up with a factor of 24 for 24 boxes. However, because of the finite size of the Spider cluster and the job queue that enables sharing of the compute resources amongst many projects competing for the same resources the actual speedup achieved was a factor \textasciitilde\,20 in real time.

\section{\texttt{DDF-Pipeline} versus recalibration for A399-401}\label{ddfvsnew}

\begin{figure*}[ht]
  \centering
  \includegraphics[width=0.8\textwidth]{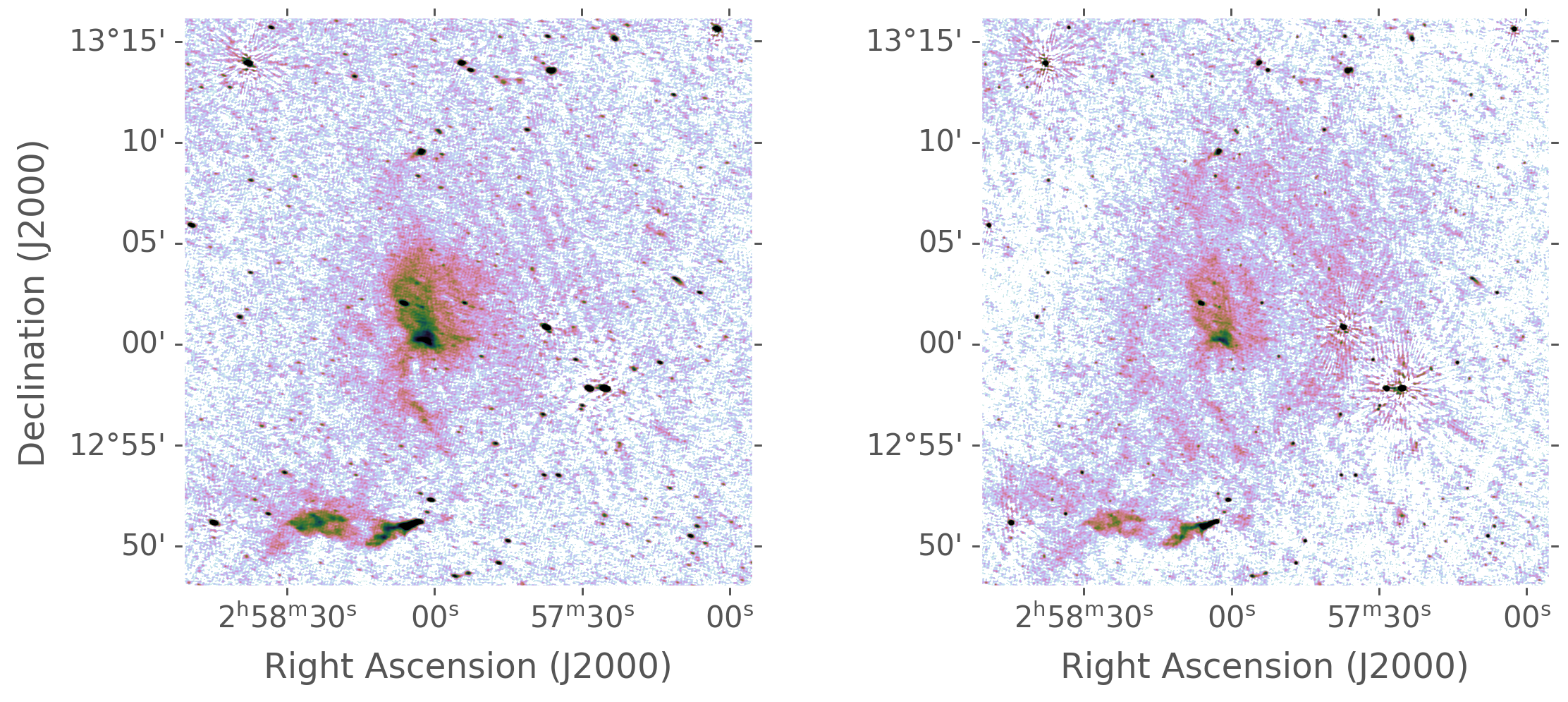}
  \centering
  \includegraphics[width=0.8\textwidth]{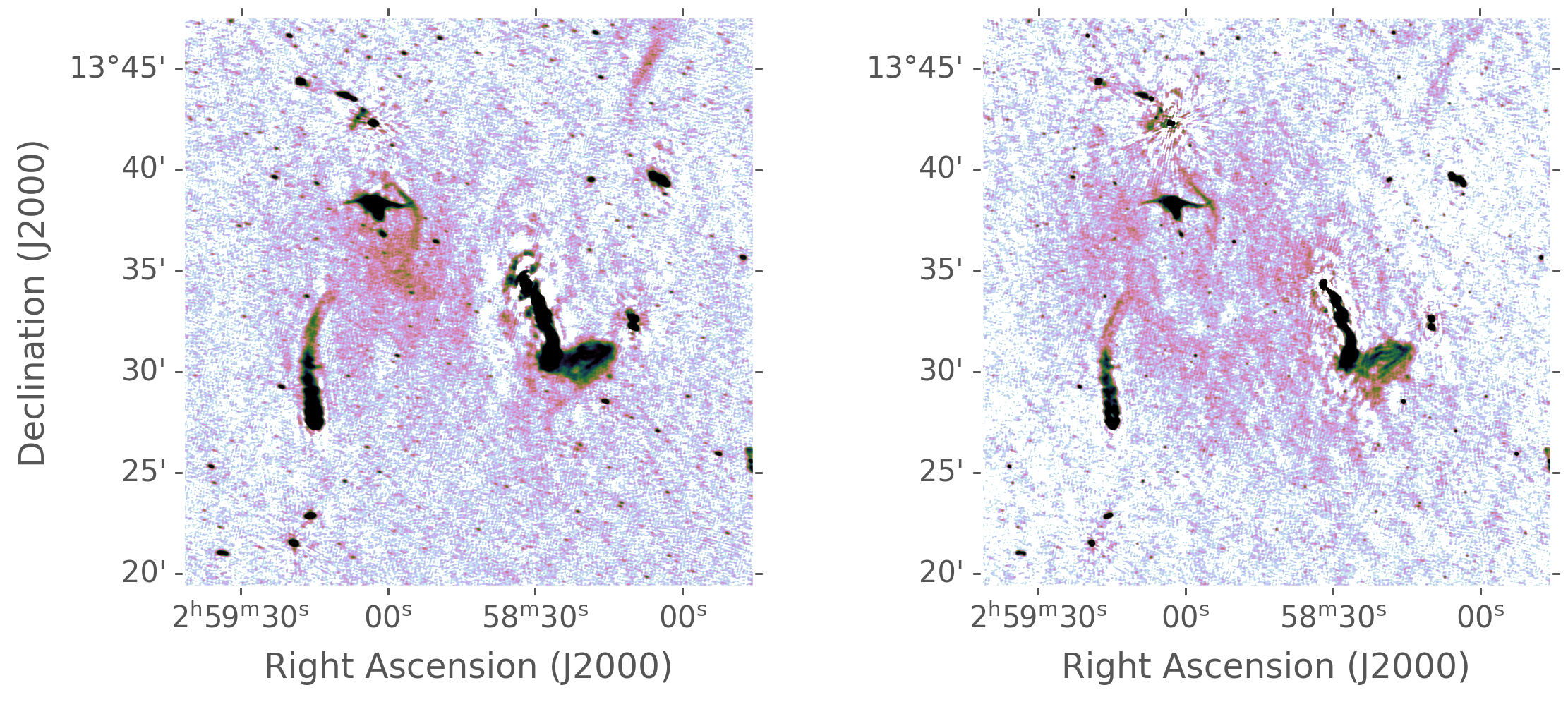}
  \centering
  \includegraphics[width=0.8\textwidth]{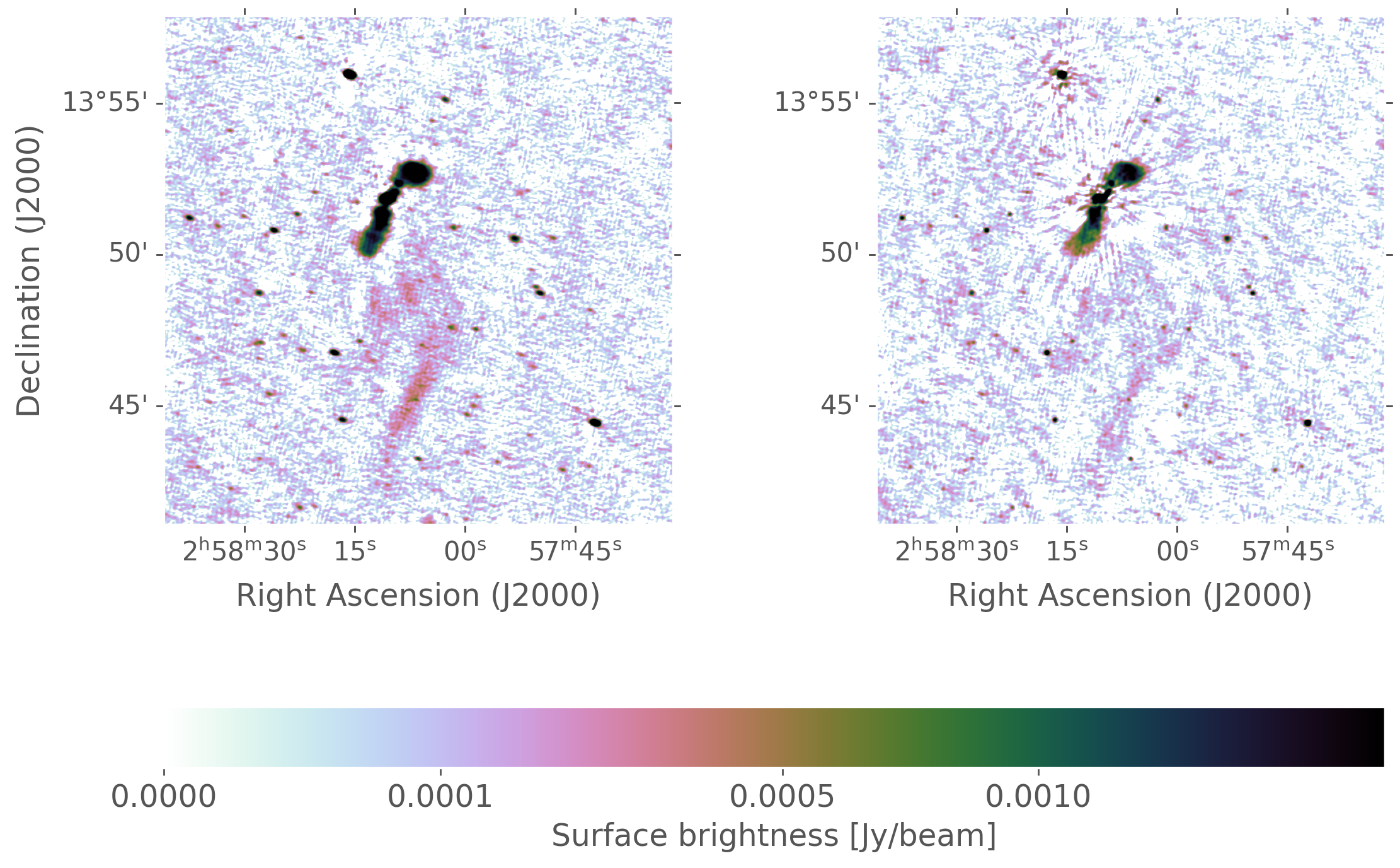} 
\caption{Image comparison between recalibration and the \texttt{DDF-Pipeline}. \textit{Left column}: Images recalibrated with the method from this paper (see Section \ref{section:calibrationmethod}). \textit{Right column}: Same images, produced with the \texttt{DDF-Pipeline} at the same color scale. The first row shows Abell 399, the second row shows Abell 401, and the last row shows the radio galaxy 2MASX\,J02581043+1351519 with its extended diffuse emission tail. The square-root scaled color bar extends from 0 to $25\sigma$ on average (average $\sigma$ from both maps).}
\label{fig:comparison}
\end{figure*}

In Section \ref{section:calibrationmethod} we described why we decided to use a more expensive calibration method than the automated \texttt{DDF-Pipeline}, which is being used by the LoTSS pipeline \citep{shimwell2019, tasse2021, shimwell2022}. We compare the final image from the \texttt{DDF-Pipeline} with our final recalibrated image that was produced with the same observations. This is not an entirely fair one-to-one comparison, as the weighting scheme in the two methods is different because of the Briggs weighting implementation in \texttt{WSClean} and the \texttt{DDF-Pipeline} image is made with its standard 100\,m baseline \textit{uv}-cut. With \texttt{WSClean,} we obtain a resolution of $5.9\arcsec\times10.5\arcsec$, while the \texttt{DDF-Pipeline} uses \texttt{DDFacet} and has a resolution of $6\arcsec\times6\arcsec$. The resulting noise levels are similar with our $\sigma=79$ $\mu\text{Jy beam}^{-1}$ versus \texttt{DDF-Pipeline} with a lower $\sigma=72$ $\mu\text{Jy beam}^{-1}$. The individual images in Figure \ref{fig:comparison} show that artifacts around bright compact sources are better suppressed with our calibration method in most cases, and diffuse emission is better reconstructed (a clear exception from the improvement is the right tailed source in the middle panel). To quantify the artifact reduction, we also studied several bright compact sources and found an improvement in a large dynamic range $\left(\frac{\text{pixel}_{\text{max}}}{|\text{pixel}_{\text{min}}|}\right)$  in most cases, by a factor of \textasciitilde1.6 on average. 

\section{\texttt{Halo-FDCA} results}\label{appendix:halofdcaresults}

\begin{figure*}[ht]
  \centering
  \includegraphics[width=1.05\textwidth]{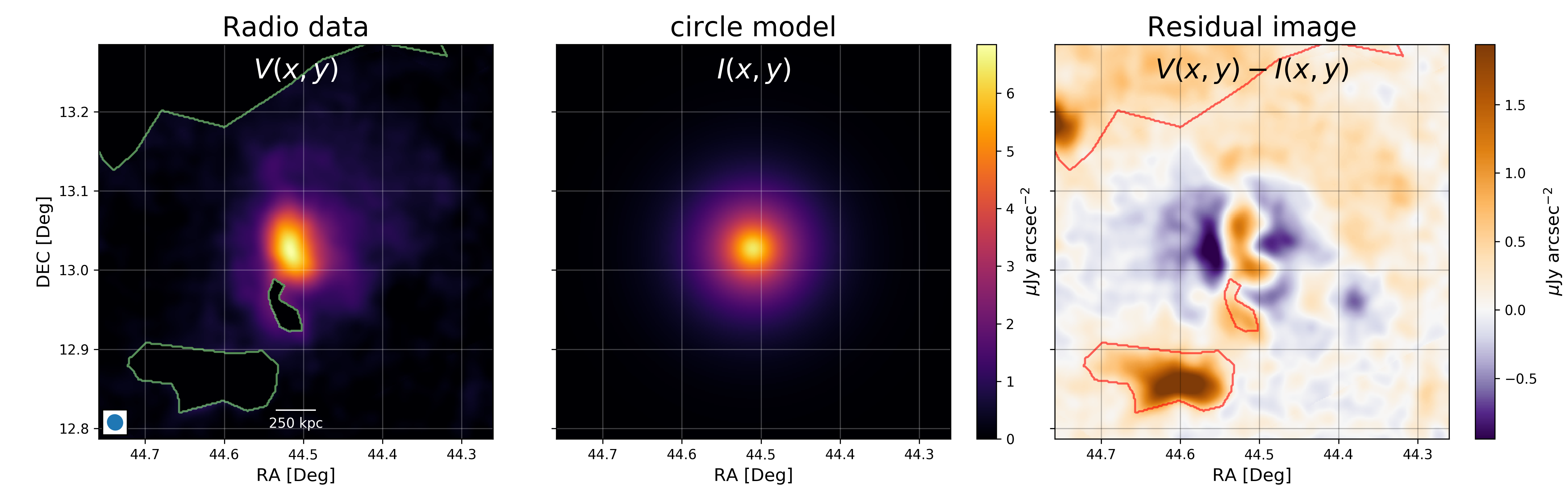}
  \centering
  \includegraphics[width=0.95\textwidth]{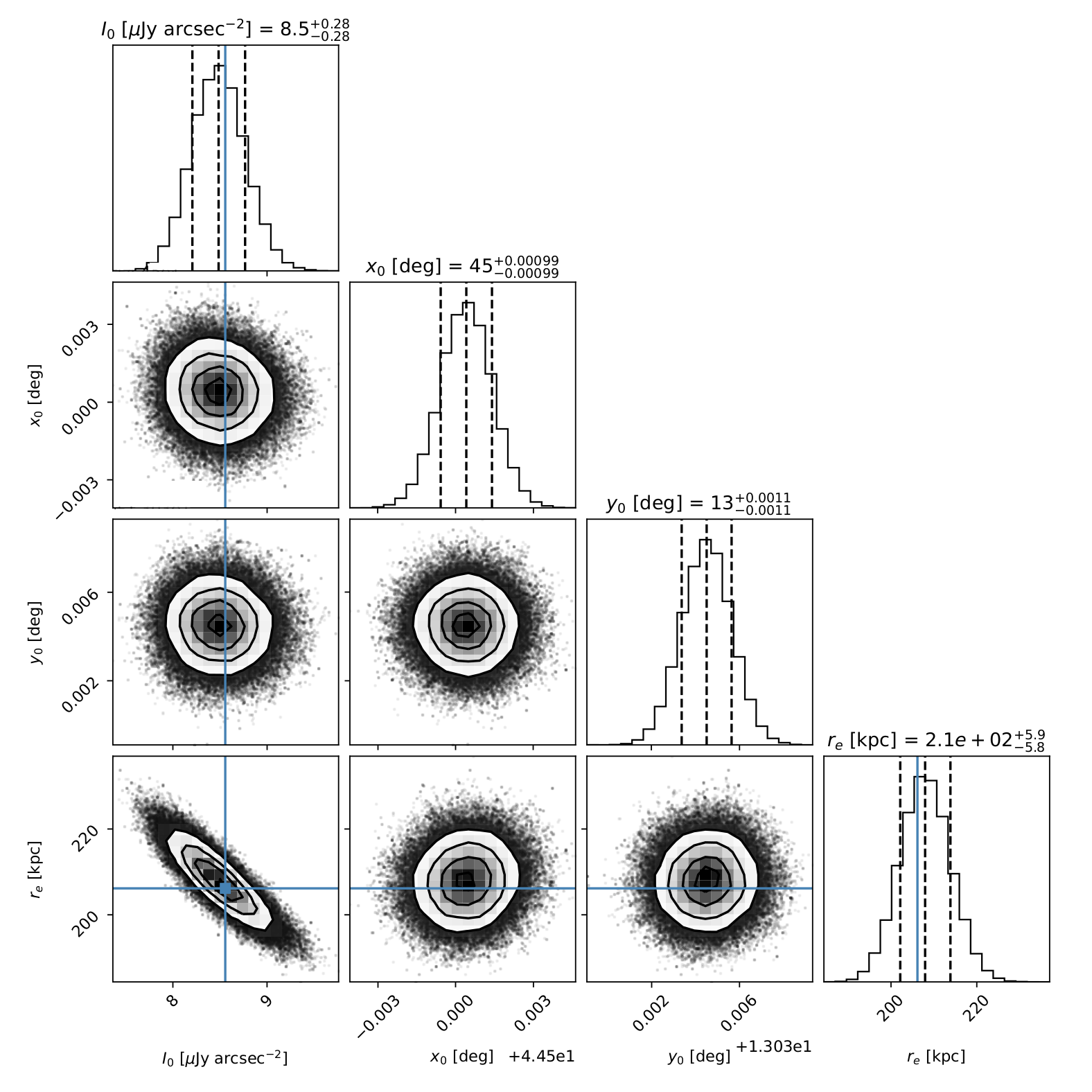}
\caption{Results obtained from fitting the radio halo in A399 with \texttt{Halo-FDCA} \citep{boxelaar2021}. \textit{Top panel}: Image for the overlay fit with corresponding masks on bright AGNs. \textit{Lower panels}: Markov chain Monte Carlo corner plot with the distributions of the posteriors of each fitted parameter.}
\end{figure*}

\begin{figure*}[ht]
  \centering
  \includegraphics[width=1.05\textwidth]{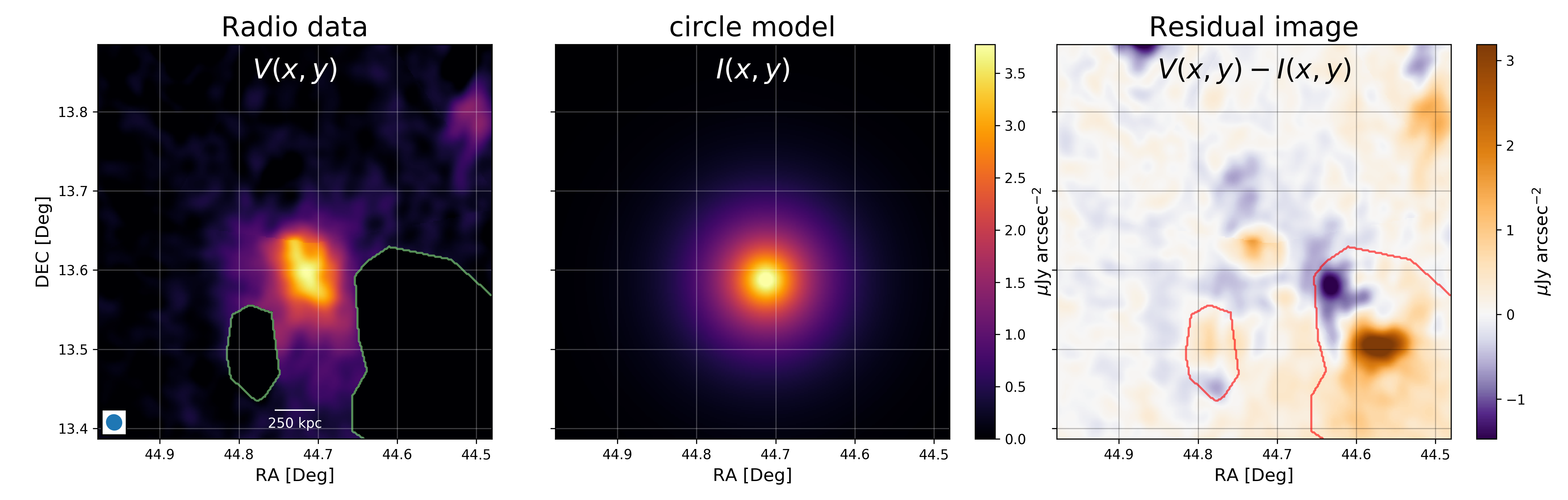}
  \centering
  \includegraphics[width=0.95\textwidth]{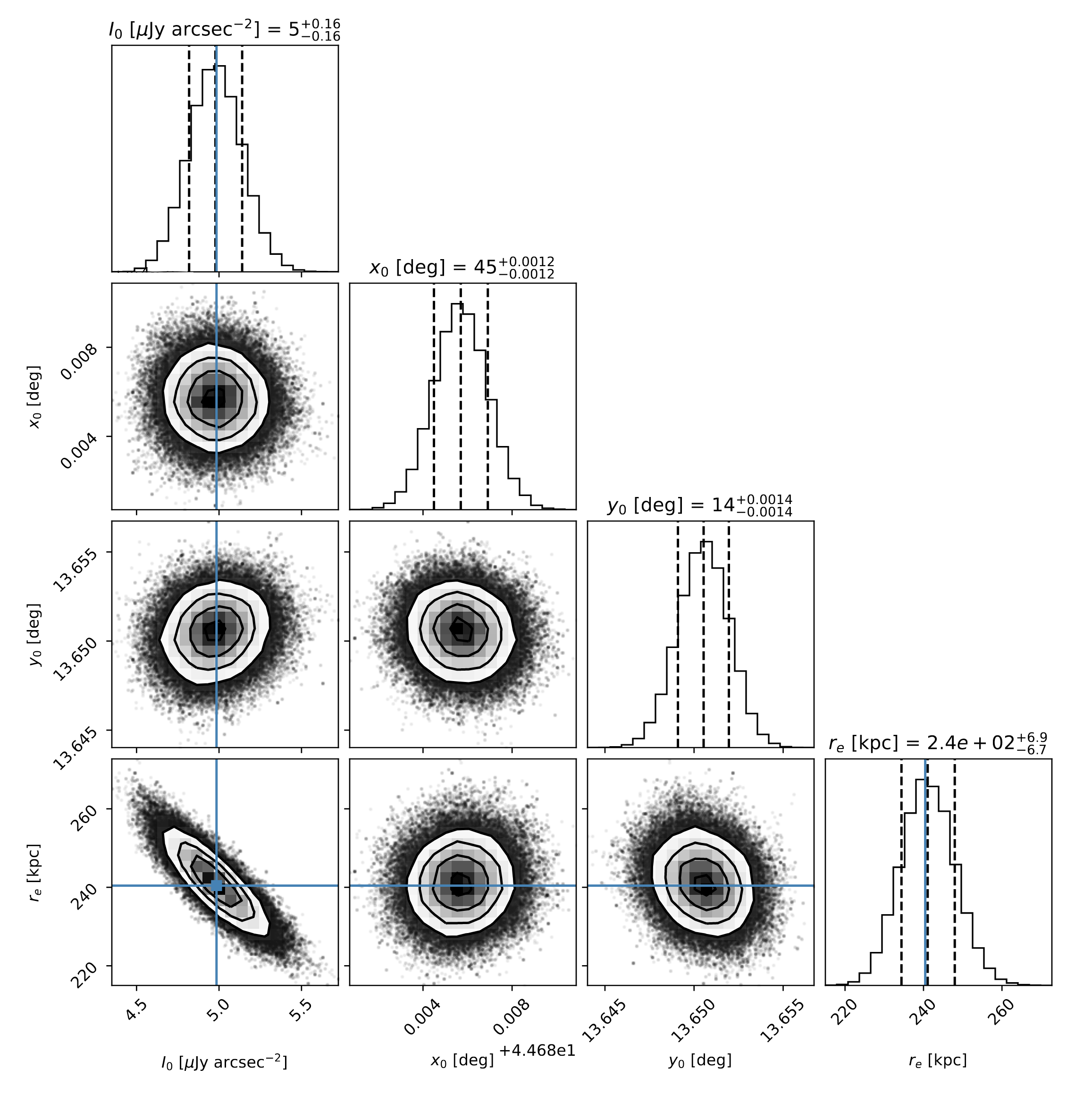}
\caption{Results obtained from fitting the radio halo in A401 with \texttt{Halo-FDCA} \citep{boxelaar2021}. \textit{Top panel}: Image for the overlay fit with corresponding masks on bright AGNs. \textit{Lower panels}: Markov chain Monte Carlo corner plot with the distributions of the posteriors of each fitted parameter.}
\end{figure*}

\end{document}